%% file: main.tex
\documentclass[5p, longtitle, authoryear]{elsarticle}

\usepackage{amssymb}
\usepackage{amsmath}

\usepackage{enumitem}
\usepackage{lineno}
\usepackage{booktabs}
\usepackage{fix-cm}
\usepackage{graphicx}
\usepackage{hyperref}
\usepackage{lipsum}
\usepackage{lscape}
\usepackage{natbib}
\usepackage{placeins}
\usepackage{multirow}
\usepackage{tabularx}
\usepackage{rotating}
\journal{Journal of High Energy Astrophysics}

\begin{document}

\begin{frontmatter}

\title{Long-term study of the gamma-ray emission of Cygnus~X-3 with MAGIC and \textit{Fermi}--LAT}
\input{author_list_20260325}

\begin{abstract}
Cygnus~X-3 is a microquasar composed of a compact object of unknown nature closely orbiting around a Wolf–Rayet star. The particularities of this source make it a unique case among microquasars. This fact, together with its recent establishment as a PeV particle accelerator, makes Cygnus~X-3 a very interesting target for the investigation of the physical processes leading to gamma-ray production. In this work, the TeV and GeV gamma-ray emission of Cygnus~X-3 is studied in order to determine its origin and constrain the properties of the system. For that purpose, a point-like analysis of 130~h of data taken with the MAGIC telescopes between 2013 and 2024 was performed, which represents the largest available sample for Cygnus~X-3 at $\sim$TeV energies. Additionally, contemporary data from \textit{Fermi}--LAT were also analysed to better contextualize the MAGIC observations. For a more detailed investigation of the source physics, the data were divided into three subsets according to the flaring state of the source and orbital phase. No significant detection of Cygnus~X-3 is found between 0.1 and 7~TeV for any of the datasets, and differential and integral flux upper limits are reported over the long-term monitoring of the source. The \textit{Fermi}--LAT fluxes can be considered compatible with previous results, taking into account the different data samples used across studies. The MAGIC upper limits presented in this work represent the most constraining ones up to date at $\sim$TeV energies. An eventual detection of Cygnus~X-3 at these energies would significantly constrain the source properties, and is not unreasonable to expect given that the source has already been detected in both the GeV and PeV regimes during flaring states. Further observations of Cygnus~X-3 at energies above tens of GeV would be valuable for this purpose.
\end{abstract}

\begin{highlights}
\item We present the largest very-high-energy data sample of Cygnus X-3 up to date.
\item The source is not detected regardless of its flaring state or orbital phase.
\item The very-high-energy flux upper limits are compatible with the GeV and TeV emission.
\item An eventual detection would shed light on the origin of this potential emission.
\end{highlights}


\end{frontmatter}

\section{Introduction}

Microquasars (MQs) are binary systems consisting of a compact object (CO) accreting material from a companion star and producing relativistic jets \citep{Mirabel1994}. The jets are launched in the vicinity of the CO and can accelerate particles up to relativistic energies \citep[e.g.,][]{Romero2017}, which in turn emit radiation from radio to multi-TeV energies \citep{Mirabel1994,LHAASO2025}. Based on their multiwavelength emission (radio and X-rays, in particular), MQs can typically be classified according to different accretion states \citep{Homan2005}. In the soft state, the thermal emission from the inner accretion disk dominates the X-ray spectra \citep[e.g.,][]{McClintock2006}, while the hard state is associated with the presence of a non-thermal X-ray component and steady radio jets \citep{Remillard2006}. Discrete relativistic ejections, occasionally resolved from their radio emission, are associated with the transition between the hard and soft states \citep{Fender2004,Fender2009}. The different accretion states and transitions between them make MQs perfect laboratories for studying accretion processes and the physics involved. In addition, the recent detection of ultra-high-energy (UHE; E > 100 TeV) gamma rays from MQs \citep{LHAASO2025, V4641Sgr} shows that accreting COs and their environments can operate as extremely efficient particle accelerators up to and above 1~PeV.

Cygnus~X-3 (RA = $20^{\rm h} 32^{\rm m} 25.78{\rm s}$, Dec = $+40^\circ 57^{\prime }27.9^{\prime\prime}$) is a MQ located at a distance of $9.7\pm 0.5$~kpc \citep{Reid2023}. The large distance to the source, together with its location near the Galactic plane, results in a high absorption along the line of sight, preventing the detection of an optical counterpart. Cygnus~X-3 has an orbital period of 4.8~h \citep[e.g.,][]{Antokhin2019}, which implies a very close orbit for the binary system. The nature of the CO is still unknown, since its derived mass range is compatible with both a black hole and a neutron star \citep[e.g.,][]{Zdziarski2013}. The companion star is a Wolf-Rayet with a mass between 8 and 14~M$_{\odot}$ \citep{vanKerkwijk1992,Koljonen2017}, which makes Cygnus~X-3 the only known MQ with a stellar companion of this type \citep{vanHeuvel2019}. Cygnus~X-3 occasionally undergoes strong radio outbursts, with fluxes reaching up to tens of Jy, originating from expanding, collimated jet-like structures \citep[e.g.,][]{Marti1992, Marti2001, Egron2017}. These events, which make Cygnus~X-3 the brightest MQ at radio frequencies, are preceded by the so-called ultrasoft state, characterized by a very high soft X-ray flux and very low or undetectable radio and hard X-ray fluxes \citep[e.g.,][]{Koljonen2018}. Contrary to what is observed in most MQs, powerful jets can be detected during the soft state, typically around 50 days after the transition from the hard state \citep{Cao2020}. As a consequence, the correlation between radio and X-rays also changes with respect to the one seen in the rest of the MQs, with radio flares taking place in soft X-ray states \citep[e.g.,][]{Koljonen2010}. These differences may be caused by the complex interaction between the jet and the strong wind from the Wolf-Rayet star \citep{Cao2020}, which takes place in the context of a very close orbit. This atypical behaviour makes Cygnus~X-3 a unique case among the MQs, and has attracted considerable scientific interest since its discovery.

High-energy (HE; $E > 100$~MeV) gamma-ray emission up to tens of GeV is detected from Cygnus~X-3 during flaring states \citep{AGILE2009,Fermi2009}, typically observed when the system is moving into or out of the ultrasoft state \citep{Corbel2012, agile2012, Zdziarski2018}. There are two main mechanisms that can explain this HE emission. In a leptonic scenario, gamma-rays are produced via inverse Compton (IC) interactions between the photons of the Wolf-Rayet star and the energetic electrons in the jet \citep[e.g.,][]{Dubus2010}. In the hadronic case, the HE emission is the result of the decay of pions, which are produced by the collisions between protons in the jets and the stellar wind (e.g., \citealt{Romero2003}; see also \citealt{Bosch2009} for a review on the non-thermal emission mechanisms in MQs). The HE emission from Cygnus~X-3 is temporally coincident with a high flux of soft X-rays, a low flux of hard X-rays, and the presence of significant radio emission with rapid variation from active relativistic jets \citep{Corbel2012}. The correlation between the emission at HE and radio suggests that the same population of electrons in the jets is responsible for the emission in both energy ranges through IC and synchrotron processes, respectively. This is reinforced by the fact that the HE flux is strongly modulated by the orbital phase, with maxima occurring around the superior conjunction of the CO \citep{Zdziarski2018}, as expected from the angular effects of the IC interactions. Therefore, the leptonic scenario is the favoured explanation to (most of) the HE emission of Cygnus~X-3.

Very-high-energy (VHE; $E > 100$~GeV) gamma-ray observations of Cygnus~X-3 were also conducted with MAGIC \citep{Aleksic2010} and VERITAS \citep{Archambault2013}, although no significant emission was detected. The close orbit of the binary system, together with the high luminosity of the companion star, provides in principle ideal conditions for the IC emission of VHE gamma rays, which contrast with the non-detection of the source to date. A possible explanation to this fact is that the same system properties that enhance the IC mechanism also increase the amount of gamma-ray absorption by pair creation with the stellar photons \citep[gamma-gamma absorption; e.g.,][]{Gould1967}. If the VHE emission was produced close enough to the base of the jet (and therefore the star), it would be highly absorbed and unable to reach an observer on Earth.

Cygnus~X-3 has been recently detected at UHE by LHAASO \citep{LHAASO2026CX3}, confirming the previously seen PeV excesses coincident with the source position obtained from the analysis of a much larger region \citep[the so-called Cygnus Bubble;][]{LHAASOcocoon2024}. In particular, Cygnus~X-3 was significantly detected for gamma-ray energies going from 60~TeV to 3.7~PeV, which establishes the source as the origin of the most energetic photons ever recorded by LHAASO, and an extreme accelerator of particles above 10~PeV. Notably, the significance of the detection mostly came from the periods in which increased HE emission was observed, whereas only flux upper limits (ULs) were reported for the quiescent HE states. Additionally, a $\sim 3\sigma$ hint of orbital modulation of the UHE emission correlated with the HE one was also observed during flaring states. This modulation, together with the monthly variability of the UHE emission among different HE states, necessarily places the UHE emitter close to the binary system, contrary to what is observed in the other MQs detected at these energies \citep{LHAASO2025}. The preferred explanation for the observed emission is the photo-meson production in inelastic collisions between relativistic protons accelerated in the inner jet and the ultraviolet and X-ray photons emitted by the Wolf-Rayet star and the accretion disk \citep{LHAASO2026CX3}.

In this work, we present the results of 12 years of monitoring of Cygnus~X-3 with the MAGIC telescopes, together with contemporary data from \textit{Fermi}--LAT. The paper is organized as follows: In Sec.~\ref{sec:obs}, the observations of Cygnus~X-3 performed with MAGIC and \textit{Fermi}--LAT are described, as well as the methods used to analyse the data. Section~\ref{sec:datasets} is devoted to explain the physical criteria used to divide the observations into different datasets. The main results from the analysis are shown in Sec.~\ref{sec:res}, including also the multiwavelength evolution of the source, while a discussion about the gamma-ray emission of Cygnus~X-3 is given in Sec.~\ref{sec:discussion}. Finally, a summary and the main conclusions of the work are given in Sec.~\ref{sec:conclusions}.

\section{Observations and data analysis}\label{sec:obs}

In this section, we briefly describe the MAGIC and \textit{Fermi}--LAT instruments, the observations performed with each of them, and the corresponding data analysis.

\subsection{VHE gamma rays (MAGIC)}\label{sec:obs:magic}

The Major Atmospheric Gamma-ray Imaging Cherenkov (MAGIC) telescopes \citep{Aleksic2016} are a system of two imaging atmospheric Cherenkov telescopes located at the Roque de los Muchachos Observatory ($28^\circ 45^\prime 22^{\prime\prime}$N, $17^\circ 53^\prime 30^{\prime\prime}$W, 2200~m above sea level), on the island of La Palma, Spain. Both telescopes have a diameter of 17 meters and are equipped with photomultiplier detectors with a field of view of $\sim 3.5^\circ$. The telescopes are designed to detect Cherenkov light from electromagnetic showers produced in the atmosphere. In their standard trigger configuration, they are sensitive to gamma rays with energies ranging from $\sim 50$~GeV up to $\sim 50$~TeV.

MAGIC has been observing Cygnus~X-3 since the start of its operations in 2006. Until 2010, observations were performed with a single telescope, while stereoscopic observations have been conducted since then. In this work, we focused on the latter, in particular on data taken from November 26, 2013 (MJD~56622) to August 13, 2024 (MJD~60535), for a total of 190 hours of observations. Part of these observations were triggered based on the radio or HE activity of Cygnus~X-3, while others were done as part of a regular monitoring program of the source. They were performed in the so-called wobble mode \citep{Fomin1994}, in which the camera centre is alternated every $\sim 20$ minutes between four different positions at an offset of $0.4^\circ$ from the position of Cygnus~X-3. This method allows for a simultaneous estimation of the background from the camera regions symmetric to the source position. After applying quality cuts to reject data affected by bad weather and/or with a high night sky background (mostly due to moonlight), a total of 129 hours of observations remain, with zenith angles between $10^\circ$ and $58^\circ$. Table~\ref{tab:observations} shows the monthly distribution of the observation time after the quality cuts, together with the corresponding zenith angle ranges.

\begin{table}[!ht]
    \caption{Monthly distribution of the MAGIC observations of Cygnus~X-3 analysed for this work. The reported observation times are after the data-quality selection cuts.}
    \label{tab:observations}
    \begin{center}
        \begin{tabular}{cccc}
            \hline
            \hline
            Year                    & Month & Obs. time (h) & Zenith angles ($^\circ$)   \\
            \hline
            \multirow{1}{*}{2013}   & Nov   & 1.3           & 31 -- 49                  \\
            \hline
            \multirow{2}{*}{2014}   & May   & 6.6           & 15 -- 40                  \\                                       & Oct   & 3.3           & 11 -- 58                  \\
            \hline
            \multirow{2}{*}{2015}   & Jul   & 0.8           & 12 -- 39                  \\  
                                    & Nov   & 1.3           & 33 -- 47                  \\
            \hline
            \multirow{3}{*}{2016}   & May   & 1.0           & 19 -- 31                  \\ 
                                    & Aug   & 9.6           & 12 -- 45                  \\ 
                                    & Sep   & 42.8          & 11 -- 52                  \\
            \hline
            \multirow{2}{*}{2018}   & Jul   & 2.1           & 12 -- 37                  \\  
                                    & Aug   & 6.2           & 11 -- 22                  \\
            \hline
            \multirow{2}{*}{2019}   & Apr   & 1.0           & 38 -- 50                  \\  
                                    & Jun   & 11.3          & 11 -- 51                  \\
            \hline
            \multirow{5}{*}{2020}   & Jun   & 4.4           & 12 -- 25                  \\ 
                                    & Jul   & 5.0           & 12 -- 23                  \\ 
                                    & Aug   & 3.2           & 12 -- 30                  \\ 
                                    & Sep   & 8.0           & 12 -- 20                  \\ 
                                    & Oct   & 1.9           & 22 -- 44                  \\
            \hline
            \multirow{2}{*}{2021}   & Apr   & 1.2           & 29 -- 46                  \\ 
                                    & Jun   & 1.9           & 27 -- 50                  \\
            \hline
            \multirow{2}{*}{2024}   & Jun   & 2.0           & 23 -- 49                  \\ 
                                    & Jul   & 14.1          & 12 -- 47                  \\
            \hline
            \multicolumn{2}{c}{Total}       & 129.0         & 11 -- 58                  \\
            \hline
        \end{tabular}
    \end{center}
\end{table}

The data analysis was performed using the MAGIC Reconstruction Software \citep[\texttt{MARS};][]{Zanin2013} and following the standard procedure described in \cite{Aleksic2016b} for a point-like source. For the purpose of this work, the analysis was done individually for each observing run, which corresponds to a $\sim 20$~min observation in a given wobble position. Individual runs were later stacked according to the dataset definition described in Sec.~\ref{sec:datasets}. The significance of the source detection was calculated using the method described in \citet{LiMa1983}. To estimate the background, an exclusion mask was set around the position of the known VHE emitter TeV~J2032+4130 \citep{tevj2032}, located at $0.5^\circ$ from Cygnus~X-3 (and thus well within the MAGIC field of view). The centre of the mask was set to the fitted position of TeV~J2032+4130 in the Cygnus~X-3 data, and a circular exclusion region with a radius of $0.24^\circ$ was chosen. This corresponds to twice the standard deviation of the fitted Gaussian, so that most of the source emission is removed from the background estimation for Cygnus~X-3. Differential and integral flux ULs were computed with a 95\% confidence level following the method described in \cite{Rolke2005} and using a total systematic uncertainty of 30\% \citep{Aleksic2012}. For this computation, the source spectral shape was taken as a power law with an index of 2.6.

\subsection{HE gamma rays (\textit{Fermi}--LAT)}\label{sec:obs:fermi}

The Large Area Telescope \citep[LAT;][]{Atwood2009} is a pair-conversion detector aboard the \textit{Fermi Gamma-Ray Space Telescope}. It is composed of a calorimeter, a tracker, an anti-coincidence detector to reject the charged-particle background, and a data acquisition system with a configurable trigger. \textit{Fermi}--LAT is in a low-Earth orbit with a 90-minute period and has a field of view of 2.4~sr, which allows it to cover the whole sky in about 3~h. The instrument can detect gamma rays with energies ranging from 20~MeV to more than 300~GeV.

A \textit{Fermi}--LAT analysis pipeline was developed years ago with the aim of monitoring the HE emission from Cygnus~X-3 \citep{Zabalza2011}. This monitoring was performed in daily time bins to search for the early phases of a flare in the source and potentially trigger MAGIC observations. Approximately half of the observing time shown in Table~\ref{tab:observations} was indeed obtained following the results of this daily monitoring, especially after 2017. The other half of the data were mostly taken during an intensive monitoring campaign performed in August and September 2016, and from additional triggers based on available multiwavelength information. In this work, two different (although similar) analysis configurations were used. One of them was devoted to the computation of a daily light curve (LC), while the other one was dedicated to the obtention of spectra for different source states, as well as a phase-folded LC during flares for the study of the orbital variability. A summary of the parameters used for both types of analyses is given in Table~\ref{tab:fermi_analysis}.

\subsubsection{Daily light curve analysis}\label{sec:obs:fermi:lc}

The computation of a daily LC was done with an updated \textit{Fermi}--LAT pipeline, which uses a newer version of the \texttt{fermitools} (v2.2.0) with the P8R3\_V3 response functions. We performed an unbinned likelihood analysis \citep{Abdo2009} from 2010--01--01 (MJD~55197) to 2024--12--31 (MJD~60675) in daily time intervals centred at 00:00~UTC, i.e., going from noon of one day to noon of the next day. This analysis is based on the spectral fitting of the sources located inside a circular region of interest (ROI) with a radius of 10$^\circ$ centred on Cygnus~X-3, and in the energy range from 0.1 to 500~GeV. In order to model the ROI, we considered all the sources in the 4FGL DR4 catalogue, which was built using all the data obtained during 14 years of observations, from 2008 to 2022 \citep{Ballet2023}. Cygnus~X-3 was included in this selection, modelled as a point-like source with a log-parabolic spectrum. We also included the components for the Galactic and isotropic backgrounds given by the standard templates, \texttt{gll\_iem\_v07.fits} and \texttt{iso\_P8R3\_SOURCE\_V3\_v1.txt}, respectively\footnote{\url{https://fermi.gsfc.nasa.gov/ssc/data/access/lat/BackgroundModels.html}}. 

The data selection was performed following the recommendations of the \textit{Fermi}--LAT collaboration\footnote{\url{https://fermi.gsfc.nasa.gov/ssc/data/analysis/documentation/Cicerone/Cicerone_Data_Exploration/Data_preparation.html}} for the point-like analysis of a Galactic source. Our analysis was performed with the energy dispersion correction enabled, except for the isotropic background. Apart from two specific cases described in Sec.~\ref{sec:obs:fermi:sp}, all the spectral parameters of the bright sources in the ROI, with a Test Statistic (TS) above 2500 in the catalogue, were left free in the fitting process. Moreover, sources that do not fulfil this TS requirement, but are classified as variable in the 4FGL DR4 catalogue, also had their normalizations freed (these are sources with a variability index above 27.69, \citealt{Ballet2023}; see also \citealt{Nolan2012} for a definition of this index). Finally, the normalizations of the Galactic and isotropic backgrounds were also left free. The rest of the spectral parameters, and all the spatial ones, were fixed to the catalogue values. Additionally, sources up to 10$^\circ$ beyond the ROI (up to 20$^\circ$ from Cygnus~X-3) were also included in the model with all their parameters fixed.

\subsubsection{Spectral and orbital analysis}\label{sec:obs:fermi:sed}

For the computation of the \textit{Fermi}--LAT spectrum and orbital LC of Cygnus~X-3, the event statistics were increased by stacking the daily data according to the criteria defined in Sec.~\ref{sec:datasets}. These increased statistics allowed us to perform a binned likelihood analysis of the stacked data, which was done with \texttt{fermipy} v1.3.1. Being a binned analysis and not part of a daily monitoring pipeline, computational time limitations were not an issue and the ROI and model radii were respectively increased to $15^\circ$ and $30^\circ$ in order to account for the contribution of more sources around Cygnus~X-3. All the sources inside the new ROI and a TS above 25 in the stacked datasets were left free in the model fit. The rest of the parameters remained the same as those used for the daily LC computation.

\subsubsection{Special cases: 4FGL J2032.2+4127 and 4FGL J2028.6+4110e}\label{sec:obs:fermi:sp}

The model-fitting criteria described in Secs.~\ref{sec:obs:fermi:lc} and ~\ref{sec:obs:fermi:sed} apply to all the sources except for two, both of which are above the defined TS thresholds: 4FGL~J2032.2+4127 and 4FGL~J2028.6+4110e (the Cygnus Cocoon). The former is a point-like source at a distance of 0.50$^\circ$ from Cygnus~X-3, and is associated to a young pulsar in a binary system with a massive star \citep{Abdo2009b}. The likely VHE counterpart of this source is TeV~J2032+4130, already mentioned in Sec.~\ref{sec:obs:magic}. The Cygnus Cocoon is an extended source modelled as a Gaussian with a 2$^\circ$ standard deviation and centred at 0.74$^\circ$ from Cygnus~X-3, and corresponds to a bubble surrounding a region of massive star formation \citep{Ackermann2011}. Due to the poor angular resolution of the LAT at low energies (95\% containment radius of $\sim 10^\circ$ at 100 MeV\footnote{\url{https://s3df.slac.stanford.edu/data/fermi/groups/canda/lat_Performance.htm}}), freeing the spectral parameters of these sources may affect the results for Cygnus~X-3. During flaring periods, some of the gamma rays emitted by Cygnus~X-3 may be wrongly associated to any (or both) of these two sources, artificially increasing their modelled flux and decreasing that of Cygnus~X-3. The opposite may also happen, with photons from 4FGL~J2032.2+4127 or the Cygnus Cocoon being associated to Cygnus~X-3 and resulting in an unreal increase of the modelled emission of the latter. All this is especially relevant when considering short time bins with limited statistics, as it is the case with the daily binning used in the LC analysis. 

Unlike Cygnus~X-3, both 4FGL~J2032.2+4127 and 4FGL~J2028.6+4110e are steady HE emitters and no significant variations in their fluxes have been observed since the start of the \textit{Fermi} mission. By performing dedicated analyses, we checked that their spectral shapes in the 2019--2024 period are compatible to those of the 2011--2016 period, both of them being in turn consistent with the parameters in the 4FGL DR4 catalogue \citep[to be safe, we are avoiding here the times around the periastron passage of 4FGL~J2032.2+4127 in November 2017, although no variability was detected at HE, as reported by][]{Li2018}. For this reason, we fixed all the model parameters of 4FGL~J2032.2+4127 and 4FGL~J2028.6+4110e to their catalogue values in order to minimize their effect on the modelled emission of Cygnus~X-3.

\begin{table}[!ht]
    \caption{Summary of the parameters used for the analysis of the \textit{Fermi}--LAT data of Cygnus~X-3. Both the common configuration and specific parameter values used for the LC and spectral analyses are listed.}
    \label{tab:fermi_analysis}
    \begin{center}
    \begin{tabular}{lc}
        \hline
        \hline
        \multicolumn{2}{c}{Common configuration}                                  \\
        \hline
        Catalogue                       & 4FGL DR4                                \\
        
        Response functions              & P8R3\_V3                                \\
        Event class                     & 128                                     \\
        Event type                      & 3                                       \\
        Max. zenith angle               & $90^\circ$                               \\
        Spatial bin width               & $0.04^\circ$                             \\
        Energy bins per decade          & 10                                      \\
        Likelihood optimizer            & \texttt{NEWMINUIT}                      \\
        Galactic background             & \texttt{gll\_iem\_v07.fits}             \\
        Isotropic background            & \texttt{iso\_P8R3\_SOURCE\_V3\_v1.txt}  \\
        Free normalizations             & Variability index above 27.69           \\
        \multirow{2}{*}{Special sources}& 4FGL~J2032.2+4127                       \\
                                        & 4FGL~J2028.6+4110e                      \\
        \hline
        \multicolumn{2}{c}{Daily light curve analysis}                            \\
        \hline
        Analysis type                   & Unbinned                                \\
        ROI radius                      & $10^\circ$                               \\
        Model radius                    & $20^\circ$                               \\
        Free spectral parameters        & TS > 2500 in 4FGL DR4                   \\
        \hline
        \multicolumn{2}{c}{Spectral and orbital analysis}                \\
        \hline
        Analysis type                   & Binned                                  \\
        ROI radius                      & $15^\circ$                               \\
        Model radius                    & $30^\circ$                               \\
        Free spectral parameters        & TS > 25 in the dataset                  \\
        \hline
    \end{tabular}
    \end{center}
\end{table}

\section{Definition of the datasets}\label{sec:datasets}

To study the overall behaviour of the source in different states, the gamma-ray data of Cygnus~X-3 were divided according to the daily \textit{Fermi}--LAT emission obtained from the analysis described in Sec.~\ref{sec:obs:fermi:lc} with daily time bins centred at midnight UTC. Cygnus~X-3 was classified as flaring at HE for a given day if either of the following two criteria was met:

\begin{itemize}[leftmargin=*, labelsep=0.5em, align=left]
\item{The daily TS of the source is above 10 and its daily flux is more than 1$\sigma$ above the median flux over the entire analysed period (2010 -- 2024).}
\vskip 2mm
\item{Cygnus~X-3 is considered to be in an "active period", defined as occurring when three or more days within a week exhibit a source TS $\gtrsim 20$, and ending when the TS falls below 10 for three consecutive days. Within an active period, all days are considered as flaring, even if there are no \textit{Fermi}--LAT results for a specific day (which may happen due to problems with the analysis or no available data).}
\end{itemize}

Each MAGIC observing run was associated to a daily \textit{Fermi}--LAT flux and TS, classifying the MAGIC runs as either flaring or non-flaring at HE. Additionally, a further division of the gamma-ray data was done based on the orbital phase of the binary system. For this purpose, we used the quadratic plus sinusoidal ephemeris given by \cite{Antokhin2019}, which is consistent within a 3\% with the one previously obtained by \cite{Zdziarski2012}. In order to keep the statistics reasonably high, we only considered two orbital phase ($\Phi$) bins, one centred at the inferior conjunction (INFC) of the CO ($0.25 \leq \Phi < 0.75$), and the other one centred at the superior conjunction (SUPC) of the CO ($0.00 \leq \Phi < 0.25$ and $0.75 \leq \Phi < 1.00$). This is a physically-motivated division, since the potential VHE emission is expected to be significantly modulated with maxima (minima) around INFC (SUPC) due to the effect of gamma-gamma absorption, contrary to the modulation observed at HE. Due to the short orbital period of the source, the division was done event-wise for \textit{Fermi}--LAT, assigning the corresponding orbital phase to each event, and run-wise for MAGIC, assigning to each $\sim 20$-min observing run the half of the orbit in which most of the time was spent.

Taking into account both divisions regarding the orbital phase and the HE flaring state, we defined three different datasets as follows (also summarized in Table~\ref{tab:datasets}):

\begin{itemize}[leftmargin=*, labelsep=0.5em, align=left]
\item{Global dataset: This dataset includes all the MAGIC observations listed in Table~\ref{tab:observations}, regardless of the source emission and orbital phase, amounting to 129~h of data. For \textit{Fermi}--LAT, it consists of all the daily time intervals in which MAGIC observations were performed, for a total of 67 daily bins of HE data.}
\vskip 2mm
\item{INFC flaring dataset: This dataset includes those MAGIC observations performed when the source was flaring at HE and the CO was in the INFC half of the orbit, and amount to 24.6~h of data. These correspond to 30 daily \textit{Fermi}--LAT bins, from which around half of the effective observation time remains after considering only events in the INFC half of the orbit.}
\vskip 2mm
\item{SUPC flaring dataset: This is the same as the INFC dataset, but in the SUPC half of the orbit. It comprises 28.9~h of MAGIC data and 30~days of \textit{Fermi}--LAT data, which are also cut in half once only the events around the SUPC are considered.}
\end{itemize}

\begin{table}
    \addtolength{\tabcolsep}{-4pt}
    \footnotesize
    \caption{Summary of the defined gamma-ray datasets. The table shows, from left to right, the dataset name, the orbital phase range, whether it considers only observations when the source was flaring at HE or not, and the total observation times with MAGIC and \textit{Fermi}--LAT.}
    \label{tab:datasets}
    \begin{center}
    \begin{tabularx}{\linewidth}{ccccc}
    \hline
    \hline
    Dataset               & $\Phi$       & HE flare             & MAGIC time (h)        & LAT time (days)      \\
    \hline
    Global                & 0.00 -- 1.00 & Indifferent          & 129.0                 & 67                   \\
    \\
    INFC                  & 0.25 -- 0.75 & Yes                  & 24.6                  & 30*                  \\
    \\
    \multirow{2}{*}{SUPC} & 0.00 -- 0.25 & \multirow{2}{*}{Yes} & \multirow{2}{*}{28.9} & \multirow{2}{*}{30*} \\
                          & 0.75 -- 1.00 &                      &                       &                      \\
    \hline
    \end{tabularx}
    \end{center}
    \footnotesize{*Approximately one half of the events of each 30-day dataset are considered due to the cut in orbital phase.}
\end{table}

For MAGIC, the results for each dataset were obtained by stacking the corresponding observation runs and performing the necessary steps to generate high-level scientific products \citep[see][]{Aleksic2016b}. For \textit{Fermi}--LAT, a \texttt{PULSE\_PHASE} column was added to the photon files with the value of the orbital phase corresponding to each event, so that an event-based stacking of the data could be applied (similarly as what is done for pulsar analyses).

\section{Results}\label{sec:res}

In this section, we present the MAGIC and \textit{Fermi}--LAT results obtained for the three datasets described in Sec.~\ref{sec:datasets}. The VHE observations of Cygnus~X-3 studied in this work show no significant point-like emission for any of the datasets and, therefore, $95\%$ confidence level ULs are reported. The obtained global integral flux UL at $E>210$~GeV is $1.68 \times 10^{-12}$~cm$^{-2}$\,s$^{-1}$.

\begin{sidewaysfigure*}[hp!]
    \centering
    \includegraphics[width=\textheight]{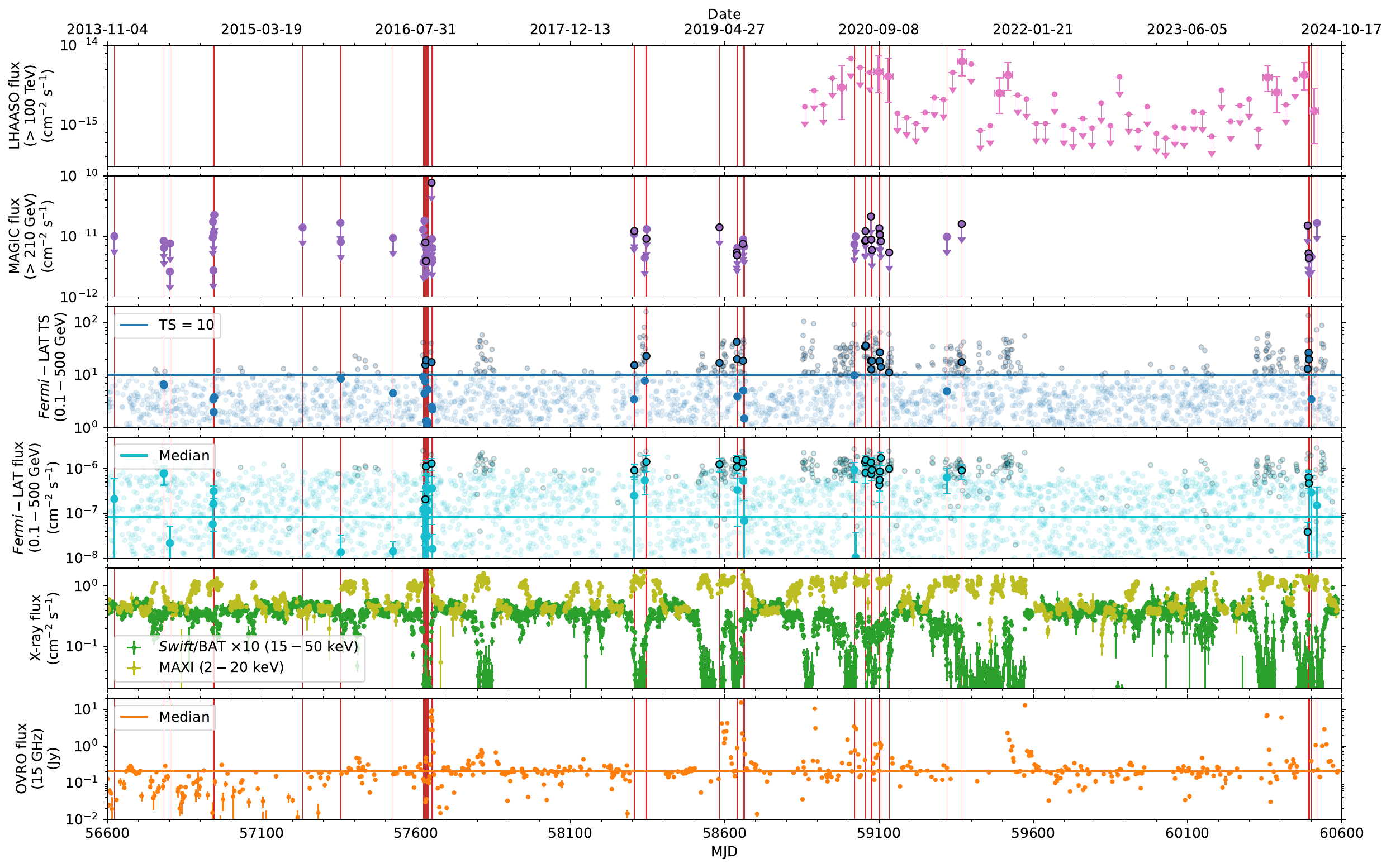}
    \caption{Multiwavelength light curve of Cygnus~X-3 at different energy ranges. From top to bottom, the panels show the LHAASO UHE flux above 100~TeV in bins of 30 days, the daily MAGIC VHE ULs above 210~GeV, the daily \textit{Fermi}--LAT TS and HE flux between 0.1 and 500~GeV, the daily X-ray fluxes from \textit{Swift}/BAT and MAXI in the 15--50~keV and 2--20~keV energy ranges, respectively, and the daily radio flux density at 15 GHz obtained with OVRO. In the top panel, LHAASO flux points are given for bins with a TS > 4, while 95\% confidence-level ULs are plotted otherwise. A black edge surrounds the HE and VHE points corresponding to days with a \textit{Fermi}--LAT TS above 10. Red vertical lines represent the days when Cygnus~X-3 was observed by MAGIC, and for which the \textit{Fermi}--LAT points are highlighted with respect to the rest of the (shaded out) daily values. For a better visualization, the \textit{Fermi}--LAT flux error bars are only shown for the highlighted points.}
    \label{fig:LC}
\end{sidewaysfigure*}

The multiwavelength LC of Cygnus~X-3 is shown in Fig.~\ref{fig:LC}, covering a time range between MJD~56600 (2013-11-04) and 60600 (2024-10-17). The figure depicts the MAGIC ULs and the \textit{Fermi}--LAT fluxes and TS values obtained in this work, the monthly fluxes reported by LHAASO \citep{LHAASO2026CX3}, the X-ray LCs from the publicly available data of MAXI/GSC\footnote{\url{http://maxi.riken.jp/star_data/J2032+409/J2032+409.html}} \citep{maxi2009} and \textit{Swift}/BAT\footnote{\url{https://swift.gsfc.nasa.gov/results/transients/CygX-3/}} \citep{swift2013}, and the radio fluxes from OVRO \citep[private communication; see][for a description of the data reduction process]{ovro2011}. The highlighted points in the \textit{Fermi}--LAT panels correspond to days coincident with MAGIC observations, which are indicated by vertical lines. Many of these points have a TS above 10 (indicated by the black contours) and a HE flux above the median, which is consistent with the fact that a big part of the MAGIC observations were triggered by the daily \textit{Fermi}--LAT results. As already mentioned, the LHAASO detections also happened during high HE states, while ULs were obtained the rest of the time. Additionally, the LC also shows that HE flares (with high TS) mostly happen during soft X-ray states (high MAXI/GSC flux, low \textit{Swift}/BAT flux), and are associated with either intermediate radio flares with flux densities $\lesssim 1$~Jy or major outbursts reaching $\gtrsim 10$~Jy.

\begin{figure}
    \includegraphics[width=\linewidth]{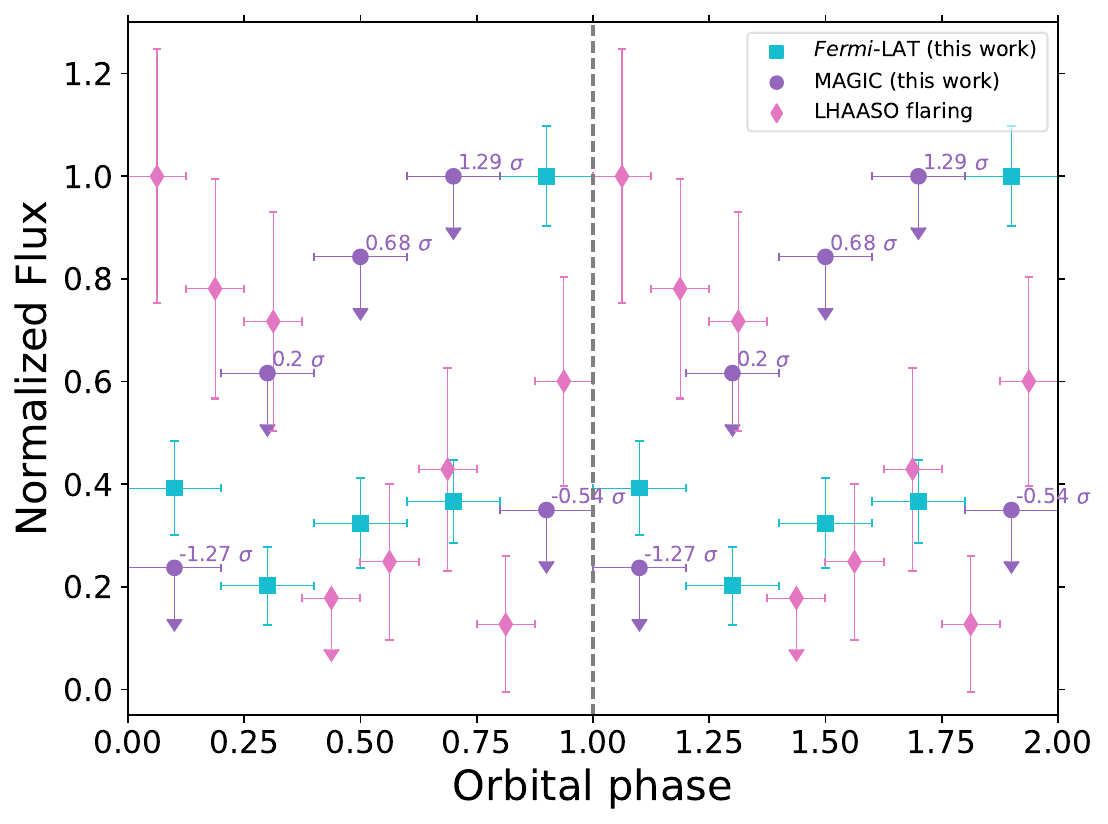} 
    \caption{Orbital LCs of Cygnus~X-3 during flaring states. Two full orbits are shown for clarity. \textit{Fermi}--LAT (0.1--500~GeV), MAGIC ($\ge210$~GeV) and LHAASO \citep[$\ge0.1$~PeV;][]{LHAASO2026CX3} data are represented as cyan squares, purple circles and pink diamonds, respectively. The detection significance in each MAGIC bin is written next to the corresponding UL. The fluxes are normalized by their maximum value for each of the LCs: $1.81 \times 10^{-6}$~cm$^{-2}$s$^{-1}$ for \textit{Fermi}--LAT, $6.11 \times 10^{-12}$~cm$^{-2}$s$^{-1}$ for MAGIC and $4.94 \times 10^{-15}$~cm$^{-2}$s$^{-1}$ for LHAASO.}
    \label{fig:Orbital}
\end{figure}

Figure~\ref{fig:Orbital} shows the normalized orbital LCs obtained from the MAGIC and \textit{Fermi}--LAT analyses during HE flaring states (i.e., for the combined INFC and SUPC datasets). The figure also depicts the LHAASO LC for data taken during \textit{Fermi}--LAT flaring states since late 2019 \citep[][we note that their criteria for defining a flare differs from ours, although we do not expect a significant change in the results due to this fact]{LHAASO2026CX3}. A clear orbital modulation is seen for both \textit{Fermi}--LAT and LHAASO, while the variations in the MAGIC ULs are compatible with statistical fluctuations.

\begin{figure}
    \includegraphics[width=\linewidth]{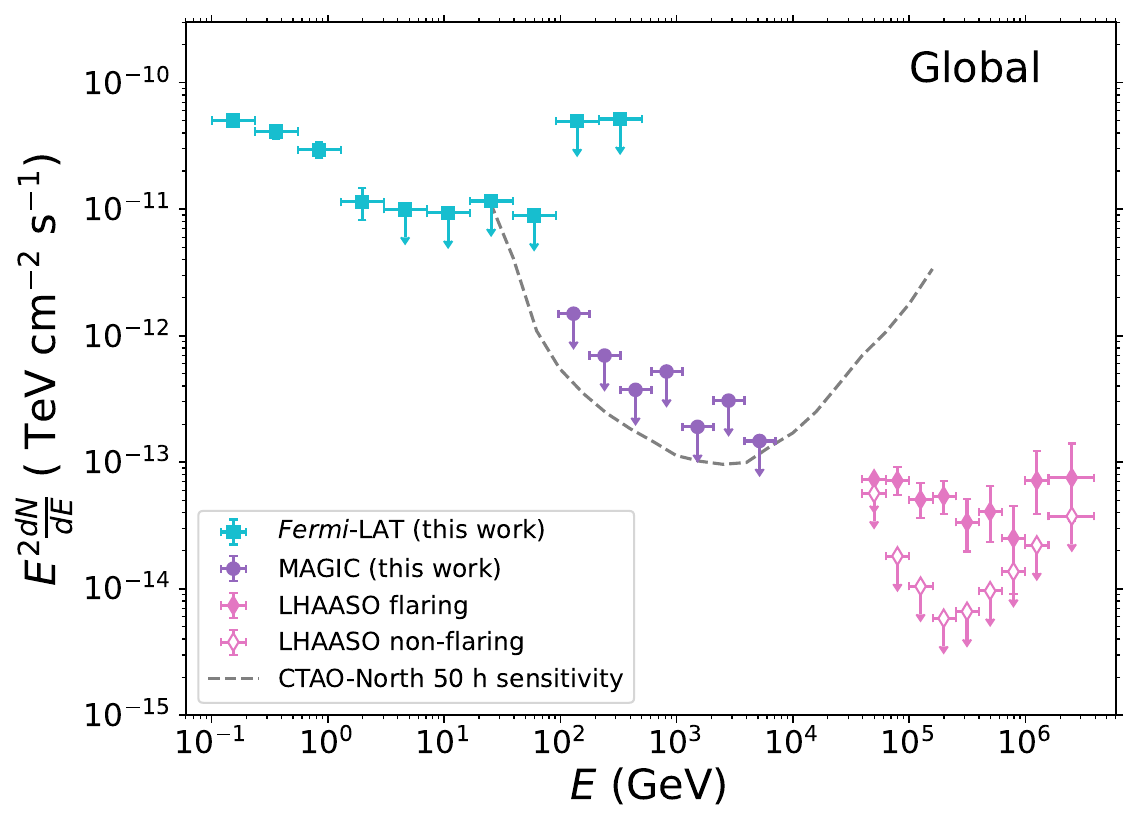}
    \vskip 2mm
    \includegraphics[width=\linewidth]{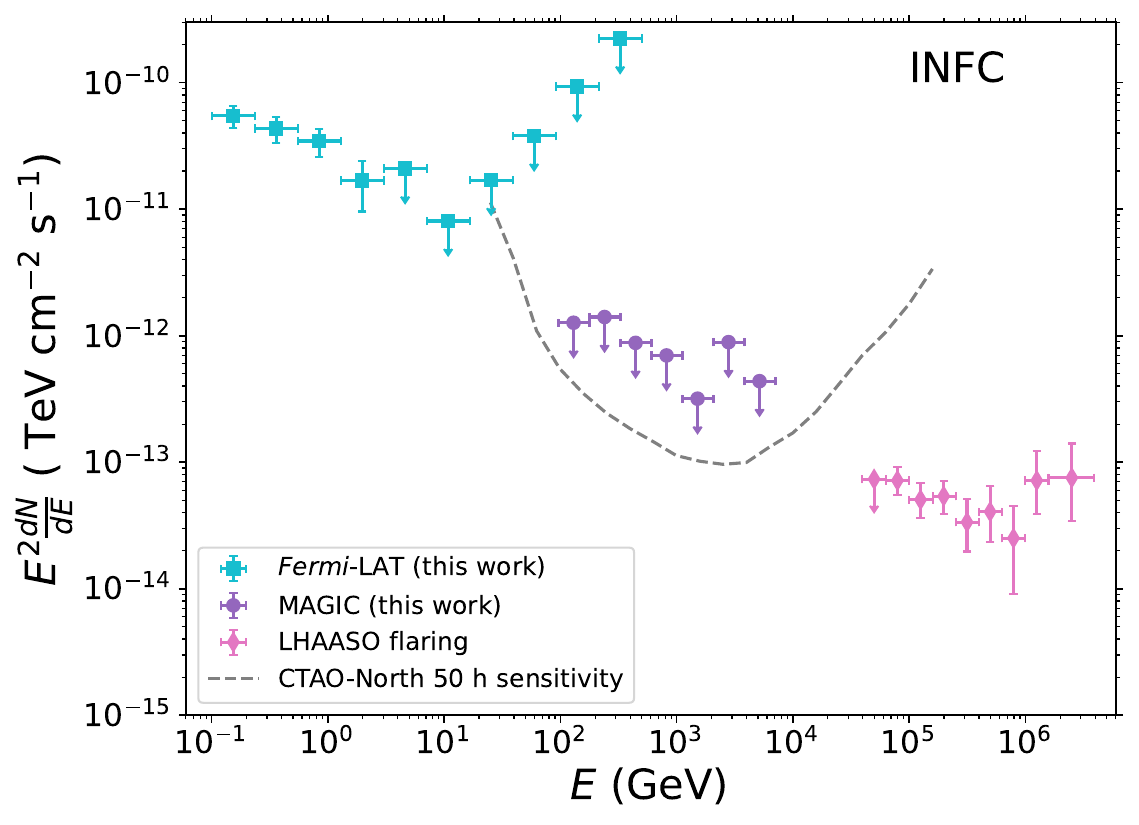}
    \vskip 2mm
    \includegraphics[width=\linewidth]{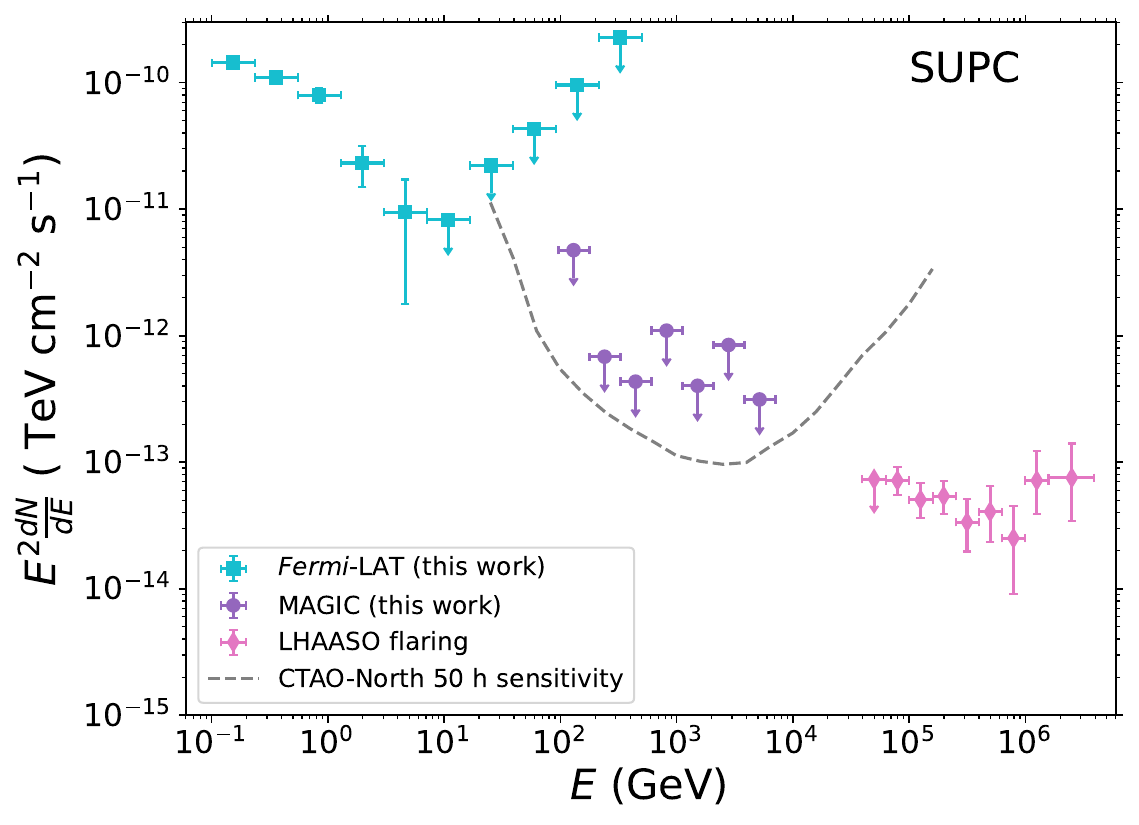}   
    \caption{Gamma-ray SEDs of Cygnus~X-3 for the Global (top panel), INFC (middle panel) and SUPC (bottom panel) datasets. MAGIC and \textit{Fermi}--LAT data are represented as purple circles and cyan squares, respectively. The top panel also shows the LHAASO SEDs from \cite{LHAASO2026CX3}, for non-flaring (empty pink diamonds) and flaring (solid pink diamonds) states. The latter is also included in the INFC and SUPC plots. The expected sensitivity for 50h of observations with the future northern array of CTAO\protect\footnotemark ~is shown as a grey dashed line.}
    \label{fig:SED}
\end{figure}

Figure~$\ref{fig:SED}$ shows the SEDs of Cygnus~X-3 for each of the three defined datasets, including both MAGIC and \textit{Fermi}--LAT data. Regarding the VHE non-detection, the flux ULs and detection significances for each energy bin are reported in Table~\ref{tab:significances:flute}. The lower energy threshold used for the Cygnus~X-3 spectrum in Fig.~$\ref{fig:SED}$ (96.4~GeV) compared to that in the LCs (210~GeV) arises from the fact that some daily bins in the latter include only data taken at moderately high zenith angles ($\gtrsim 45^\circ$). At these angles, the reconstruction process of lower-energy gamma-ray events is degraded due to the higher absorption of the Cherenkov light in the atmosphere. An increased energy threshold was therefore chosen in the LC in order to minimize the systematics in the analysis of the affected daily bins. The \textit{Fermi}--LAT fluxes obtained for the SUPC dataset are in general compatible to those reported by \cite{Zdziarski2018} and \cite{Dmytriiev2024}, although they are up to a factor of $\sim 2$ lower for the Global dataset. This discrepancy can be likely explained by the use of different data samples. While we focused only on the \textit{Fermi}--LAT data coincident with the MAGIC observations, they considered the whole available data since the start of the mission, resulting in much higher statistics. Additionally, our definition of a flaring state also changes with respect to other publications and is affected by the specific choice of some analysis parameters, in particular by how the bright sources nearby Cygnus~X-3 are treated (see Sec.~\ref{sec:obs:fermi:sp}). 

The LHAASO SEDs for both flaring and non-flaring states are also shown in Fig.~\ref{fig:Orbital}. A gap of an order of magnitude in energy is present between the last MAGIC bin and the first LHAASO spectral point at $\sim 60$~TeV for the flaring case. Nonetheless, even a pure power-law extrapolation of the LHAASO flux (highly unlikely in reality) would fall below the MAGIC ULs, a fact that does not allow the VHE observations to constrain the physical processes responsible for the UHE emission. For completeness, we also analysed the subsample of MAGIC data contemporary to the LHAASO observations during flaring states, which correspond to the 41.7~h taken since 2020. No significant differences were found between the behaviour of these data and the overall results using data taken since 2013 (shown in Fig~\ref{fig:SED}).

\addtocounter{footnote}{-1}
\footnotetext{{\url{https://www.ctao.org/for-scientists/performance/}}}

\begin{table}
    \addtolength{\tabcolsep}{-2pt}
    \footnotesize
    \caption{Flux ULs and significances derived from the MAGIC observations of Cygnus~X-3 for each energy bin and the three datasets defined in Sec.~\ref{sec:datasets}.}
    \label{tab:significances:flute}
    \begin{center}
    \begin{tabularx}{\linewidth}{cccccccc}
    \hline
    \hline
    \multicolumn{2}{c}{Energy bin}  & \multicolumn{3}{c}{SED UL} & \multicolumn{3}{c}{Significance} \\
    \multicolumn{2}{c}{(GeV)}       & \multicolumn{3}{c}{(10$^{-13}$ TeV cm$^{-2}$ s$^{-1}$)} & \multicolumn{3}{c}{($\sigma$)}\\
    \hline
    $E_{\rm min}$ & $E_{\rm max}$ & Global & INFC & SUPC & Global & INFC   & SUPC       \\
    \hline
    96.40         & 178.13        & 14.9   & 12.7 & 47.4 & 1.07    & $-0.65$ & 2.15     \\ 
    178.13        & 329.16        & 6.97   & 14.1 & 6.83 & 0.68    & 0.45    & $-0.64$  \\ 
    329.16        & 608.24        & 3.73   & 8.78 & 4.34 & 0.33    & 0.36    & $-0.52$  \\ 
    608.24        & 1123.24       & 5.21   & 6.98 & 11.0 & 1.91    & 0.71    & 2.03     \\ 
    1123.24       & 2076.88       & 1.90   & 3.18 & 4.02 & 0.05    & $-0.53$ & 0.12     \\ 
    2076.88       & 3837.75       & 3.07   & 8.90 & 8.45 & 1.03    & 1.63    & 2.00     \\ 
    3837.75       & 7091.59       & 1.47   & 4.37 & 3.14 & $-0.33$ & 0.00    & $-0.63$  \\
    \hline
    \end{tabularx}
    \end{center}
\end{table}

\section{Discussion}\label{sec:discussion}

The results of the MAGIC observations show no detection or strong hint of emission from Cygnus~X-3 regardless of the HE state or orbital phase. Nevertheless, the recent LHAASO detection of Cygnus~X-3 in gamma rays above 1~PeV during HE flares undoubtedly places this source as an extreme particle accelerator up to energies well above the MAGIC ones \citep{LHAASO2026CX3}. This fact, together with the \textit{Fermi}--LAT detection of the source, makes it reasonable to expect some degree of emission from Cygnus~X-3 at VHE. This may also be somehow suggested by the significances reported in Table~\ref{tab:significances:flute}, which in some cases go up to $\sim 2\sigma$. Therefore, albeit speculative and qualitative in nature, it may be instructive to give a discussion about the underlying physical processes that could account for potential VHE emission in the source.

An extrapolation of the \textit{Fermi}--LAT spectrum may be compatible with emission at the MAGIC lowest and mid energies during HE flares. This putative emission could be the result of IC scattering of stellar photons by the jet electrons, as in HE gamma rays \citep[e.g.,][]{Dubus2010}. Although this would require a harder electron energy distribution at higher energies due to the softening effect of the IC emissivity in the Klein-Nishina regime, this harder distribution is expected under dominant IC losses \citep{Khangulyan2014} or non-radiative cooling in a multi-zone accelerator and emitter \citep{Khangulyan2018}. The latter would be consistent with the requirement that gamma-gamma absorption in the stellar photon field should be small at the MAGIC energies if the signals are real, which would place the VHE emitter relatively far from the jet base \citep{Bednarek2010}. The low level of absorption already found at 20~GeV by \cite{Dmytriiev2024} --who used a larger set of \textit{Fermi}--LAT data-- fits in this picture.

At higher energies, if potential emission of a few TeV was present and produced by IC scattering, it would require an even harder energy distribution of electrons above $\sim 1$~TeV. An additional constraint on the stellar IC scenario for multi-TeV gamma rays is the electron maximum energy. We adopt an acceleration energy gain rate of $\dot{E}_{\rm acc} = \eta_{\rm acc}eBc$ , where $\eta_{\rm acc} \le 1$ is the acceleration efficiency, $e$ is the elementary charge and $c$ is the speed of light, and compare it with the synchrotron energy losses for a given magnetic field $B$, $\dot{E}_{\rm sync} = -1.6\times10^{-3} B^2 E^2$ (in cgs units), which are the dominant ones at the highest energies. This yields a maximum electron energy of $E_{\rm max,e} \approx 60~\eta_{\rm acc}^{1/2}\,B_{\rm G}^{-1/2}$~TeV, where $B_{\rm G}$ is the magnetic field strength in Gauss. The latter can be parametrized through the fraction $\eta_B$ of jet power in Poynting flux luminosity, which results in $E_{\rm max,e} \approx 1~\eta_{\rm acc}^{1/2}\,\eta_B^{-1/4}\,(R_{\rm j}/3\!\times\!10^{10})^{1/2}\,(L_{\rm j}/10^{38})^{-1/4}$~TeV, where $R_{\rm j}$ and $L_{\rm j}$ are the (jet) accelerator size and power in cgs units, respectively. This implies that the magnetic field should be well below equipartition ($\eta_B \ll 1$) for multi-TeV electrons to be present. In turn, a low $B$ makes a hard electron energy distribution more feasible, since the synchrotron cooling becomes less efficient. Additionally, for a star with a surface temperature of $\sim 10^5$~K and a radius of $\sim 10^{11}$~cm, the optical depth of gamma rays with energies of several TeV on the binary scales ($\sim 3\times 10^{11}$~cm) is $\tau_{\gamma\gamma}\lesssim 1$, which is significantly less than the maximum optical depth of $\tau_{\gamma\gamma}\sim 30$ reached for energies of a few tens of GeV. This represents a weakening of gamma-gamma absorption at higher gamma-ray energies, making it easier to detect multi-TeV photons originating close to the binary system and consequently hardening the observed VHE SED (to a modest degree if only a small part of the VHE photons came from the binary scales). We note that the low magnetic fraction $\eta_B$ required for potential leptonic emission at a few TeV would prevent protons from reaching the energies necessary to emit PeV gamma rays (see below). This implies that, if multi-TeV emission of leptonic origin was produced in Cygnus~X-3, it should come from a different region (with a lower $\eta_B$) than the UHE hadronic emission seen by LHAASO \citep[see also the recent study by][]{Zhang2026}.

The fact that multi-TeV emission could originate from within the binary system, due to the relatively low gamma-gamma opacities, might enable the contribution of hadronic mechanisms to the overall emission of the source. At these small scales, processes such as nuclei photo-disintegration \citep{Bednarek2005}, proton-proton interactions in the stellar wind \citep{Romero2003} or photo-meson production \citep{Levinson2001} in the X-ray photon field  may become efficient due to the higher target densities. In fact, \cite{LHAASO2026CX3} find
photo-meson interactions in the X-ray and stellar photon fields to be the likely responsible for the emission at energies around 100~TeV and 1~PeV, respectively. This emission could potentially extend to lower gamma-ray energies reachable by MAGIC for a sufficiently bright and hard intrinsic X-ray target spectrum, or for a substantial contribution from proton-proton interactions. In any case, the protons involved in hadronic processes do not have a maximum energy limited by synchrotron losses and they could be accelerated up to $E_{\rm max,p} \lesssim eBR_{\rm j} \approx 35\,\eta_B^{1/2}\,(L_{\rm j}/10^{38})^{1/2}$~PeV, with $L_{\rm j}$ in cgs units. These energies are well above the necessary ones to produce gamma rays somewhere within the MAGIC energy range via any of the hadronic processes mentioned above \citep[see the discussion in][]{Bosch2009}.

Finally, although the focus of this discussion is on the origin of a potential VHE signal, it is worth noting that several factors could contribute to its non-detection. The transient nature of Cygnus~X-3 implies that VHE emission may not be strictly simultaneous with the HE flares used to trigger MAGIC observations, potentially occurring earlier and causing part of it to remain unobserved. This effect is expected to be limited for the dataset presented in this work, as the MAGIC observations were typically carried out over several days spanning periods of a few weeks, while the source was active at HE.

Gamma-gamma absorption on the stellar photon field is expected to play a significant role in the 0.01--10~TeV regime. A potential VHE emitter located within the binary scales would experience substantial attenuation for energies below several TeV. As discussed above, the low level of absorption inferred from \textit{Fermi}--LAT observations up to 20~GeV \citep{Dmytriiev2024} suggests that the HE emitter is located at distances sufficiently far from the binary system. If this emission extends to higher energies and is produced under moderate absorption, it could reach the lower energy threshold of MAGIC. Under this scenario, the non-detection of Cygnus X-3 near the edges of the MAGIC energy range --at $\sim 100$~GeV and $\sim 10$~TeV-- may also be explained by a lack of sufficient intrinsic emission rather than by absorption effects alone.

In this context, the improved sensitivity and lower energy threshold of the future Cherenkov Telescope Array Observatory (CTAO) will provide a significant advantage. Figure~\ref{fig:SED} shows the expected sensitivity for 50~h of observations with the northern CTAO array, which lies well below the VHE ULs obtained in this work and effectively bridges the gap between the HE and VHE regimes. This enhanced performance substantially increases the chance of detecting Cygnus~X-3 at VHE. Indeed, some estimates predict that CTAO could detect the source within the 0.1--1~TeV range with exposure times ranging from a few hours to several tens of hours, depending on the assumed extrapolation of the flaring \textit{Fermi}--LAT spectrum \citep[and without accounting for gamma-gamma absorption;][]{cta2025}. These exposures could be relaxed by extending the analysis down to the CTAO-North threshold of $\sim 20$~GeV, and they could realistically be accumulated within 3--4 years, given the current recurrence rate of flares.

From all this, we can conclude that an eventual detection of the source at VHE, for which some degree of emission might be expected, would disentangle both the origin of these gamma rays and the properties of the non-thermal processes behind them. Along with this, additional information could also be inferred, such as the location of the VHE emitter and the related matter and radiation local densities (deep within or outside the binary), the nature of the emitting particles (electrons or protons/nuclei), the acceleration efficiency (moderate or very high), or the magnetic field (well below or close to equipartition).

\section{Summary and conclusions}\label{sec:conclusions}

In this work, we have analysed around 130~h of stereoscopic observations of Cygnus~X-3 taken between 2013 and 2024 with MAGIC, together with contemporary data from \textit{Fermi}--LAT. This represents the largest available VHE sample of the source, and the resulting ULs are the most constraining to date. Around half of the MAGIC observations were performed when the source was flaring in HE gamma rays according to the \textit{Fermi}--LAT daily flux monitoring. These flares mostly happened during soft X-ray spectral states and typically had an associated radio outburst, which suggests a common origin of the HE and radio emission.

The large amount of observing time devoted to the source with MAGIC allowed us to divide the data into different sets based on the Cygnus~X-3 HE gamma-ray activity and orbital phase. No VHE detection was obtained for any of the datasets, although some emission in this energy range could be expected in view of the source detection at both HE and UHE during flares. The orbital variability seen in these two energy ranges causally places the gamma-ray emitter close to the binary system, unlike what is observed in most of the other MQs detected in gamma rays. An eventual detection of Cygnus~X-3 at VHE would provide insight into the physical origin of these gamma-rays, constraining the location and properties of the VHE emitter and potentially distinguishing between the leptonic or hadronic nature of the emission. Further MAGIC observations of Cygnus~X-3 during HE flaring states are ongoing, with the aim of boosting the observed low-significance signals and determining if they are genuine. In this direction, the increased sensitivity and lower energy threshold of the future CTAO could allow for such a determination in a significantly shorter time.

\section*{CRediT authorship contribution statement}

Notable individual contributions to the paper, in alphabetical order:
\textbf{L. Barrios-Jim\'enez:} data curation, visualization, software, formal analysis, writing -- original draft, writing -- review and editing;
\textbf{V. Bosch-Ramon:} conceptualization, formal analysis, writing -- review and editing;
\textbf{M. Carretero-Castrillo:} data curation, software, validation, writing -- review and editing;
\textbf{E. Molina:} supervision, data curation, software, formal analysis, writing -- original draft, writing -- review and editing, conceptualization;
\textbf{J.M. Paredes:} conceptualization;
\textbf{M. Rib\'o:} conceptualization, writing -- review and editing.
The rest of the authors have contributed in one or several of the following ways: design, construction, maintenance and operation of the instrument(s); preparation and/or evaluation of the observation proposals; data acquisition, processing, calibration and/or reduction; production of analysis tools and/or related Monte Carlo simulations; discussion and approval of the contents of the draft.

\section*{Acknowledgements}

We would like to thank the Instituto de Astrof\'{\i}sica de Canarias for the excellent working conditions at the Observatorio del Roque de los Muchachos in La Palma. The financial support of the German BMFTR, MPG and HGF; the Italian INFN and INAF; the Swiss National Fund SNF; the grants PID2022-136828NB-C41, PID2022-137810NB-C22, PID2022-138172NB-C41, PID2022-138172NB-C42, PID2022-138172NB-C43, PID2022-139117NB-C41, PID2022-139117NB-C42, PID2022-139117NB-C43, PID2022-139117NB-C44, CNS2023-144504 funded by the Spanish MCIN/AEI/ 10.13039/501100011033 and "ERDF A way of making Europe"; the Indian Department of Atomic Energy; the Japanese ICRR, the University of Tokyo, JSPS, and MEXT; the Bulgarian Ministry of Education and Science, National RI Roadmap Project DO1-400/18.12.2020 and the Academy of Finland grant nr. 320045 is gratefully acknowledged. This work has also been supported by Centros de Excelencia ``Severo Ochoa'' y Unidades ``Mar\'{\i}a de Maeztu'' program of the Spanish MCIN/AEI/ 10.13039/501100011033 (CEX2019-000918-M, CEX2021-001131-S, CEX2024001442-S), by AST22\_00001\_9 with funding from NextGenerationEU funds and by the CERCA institution and grants 2021SGR00426, 2021SGR00607 and 2021SGR00773 of the Generalitat de Catalunya; by the Croatian Science Foundation (HrZZ) Project IP-2022-10-4595 and the University of Rijeka Project uniri-prirod-18-48; by the Deutsche Forschungsgemeinschaft (SFB1491) and by the Lamarr-Institute for Machine Learning and Artificial Intelligence; by the Polish Ministry of Science and Higher Education grant No. 2025/WK/04; by the European Union (ERC, MicroStars, 101076533); and by the Brazilian MCTIC, the CNPq Productivity Grant 309053/2022-6 and FAPERJ Grants E-26/200.532/2023 and E-26/211.342/2021.\\

This research has made use of data from the OVRO 40-m monitoring program (Richards, J. L. et al. 2011, ApJS, 194, 29). This monitoring program has been supported by NSF grants AST-0808050 and AST-1109911, and is currently supported by NSF grant AST-2407603 and AST-2407604. It has also been supported by NASA grants NNX08AW31G, NNX11A043G, and NNX14AQ89G. This research has made use of the MAXI data provided by RIKEN, JAXA and the MAXI team, and the \textit{Swift}/BAT transient monitor results provided by the \textit{Swift}/BAT team. We acknowledge the \textit{Fermi}--LAT collaboration for making available the data and the analysis tools used in this work.\\

We would also like to thank Dr.~M.~Mart\'inez-Chicharro for her valuable contribution during the early stages of the MAGIC data analysis. VB-R is Correspondent Researcher of CONICET, Argentina, at the IAR.

\section*{Data availability}

Part of the data underlying this article are available in the article and in its online supplementary material. In particular, MAXI/GSC data were obtained from \url{https://maxi.riken.jp/star_data/J2032+409/J2032+409.html}; \textit{Swift}/BAT data were downloaded from \url{https://swift.gsfc.nasa.gov/results/transients/CygX-3/}; and \textit{Fermi}--LAT data were taken from \url{https://fermi.gsfc.nasa.gov/cgi-bin/ssc/LAT/LATDataQuery.cgi}. The rest of the data will be shared on reasonable request to the corresponding authors.

\bibliographystyle{elsarticle-harv}
\bibliography{bibliography}

\end{document}

%% file: author_list_20260325.tex
%
\author[1]{K.~Abe}
\author[2]{S.~Abe}
\author[3]{J.~Abhir}
\author[4]{A.~Abhishek}
\author[5]{V.~A.~Acciari}
\author[6]{A.~Aguasca-Cabot}
\author[7]{I.~Agudo}
\author[8]{I.~Albanese}
\author[9]{T.~Aniello}
\author[10,39]{S.~Ansoldi}
% ...

\address[1]{Japanese MAGIC Group: Department of Physics, Tokai University, Hiratsuka, 259-1292 Kanagawa, Japan}
\address[2]{Japanese MAGIC Group: Department of Physics, Kyoto University, 606-8502 Kyoto, Japan}
\address[3]{ETH Z\"urich, CH-8093 Z\"urich, Switzerland}
\address[4]{Universit\`a di Siena and INFN Pisa, I-53100 Siena, Italy}
\address[5]{Institut de F\'isica d'Altes Energies (IFAE), The Barcelona Institute of Science and Technology (BIST), E-08193 Bellaterra (Barcelona), Spain}
\address[6]{Universitat de Barcelona, ICCUB, IEEC-UB, E-08028 Barcelona, Spain}
\address[7]{Instituto de Astrof\'isica de Andaluc\'ia-CSIC, Glorieta de la Astronom\'ia s/n, 18008, Granada, Spain}
\address[8]{Universit\`a di Padova and INFN, I-35131 Padova, Italy}
\address[9]{National Institute for Astrophysics (INAF), I-00136 Rome, Italy}
\address[10]{Universit\`a di Udine and INFN Trieste, I-33100 Udine, Italy}
% ...
\address[39]{also at International Center for Relativistic Astrophysics (ICRA), Rome, Italy}
% ...

\cortext[cor1]{Corresponding authors: L. Barrios-Jim\'enez, M. Carretero-Castrillo, E. Molina (Email: \texttt{contact.magic@mpp.mpg.de})}

%% file: bibliography.bib
@ARTICLE{Aleksic2016,
       author = {{Aleksi{\'c}}, J. and others},
        title = "{The major upgrade of the MAGIC telescopes, Part I: The hardware improvements and the commissioning of the system}",
      journal = {Astroparticle Physics},
         year = 2016,
        month = jan,
       volume = {72},
        pages = {61-75},
          doi = {10.1016/j.astropartphys.2015.04.004},
archivePrefix = {arXiv},
       eprint = {1409.6073},
 primaryClass = {astro-ph.IM},
       adsurl = {https://ui.adsabs.harvard.edu/abs/2016APh....72...61A}
}

@INPROCEEDINGS{Zanin2013,
       author = {{Zanin}, R. and others},
        title = "{MARS, The MAGIC Analysis and Reconstruction Software}",
    booktitle = {International Cosmic Ray Conference},
         year = 2013,
       series = {International Cosmic Ray Conference},
       volume = {33},
        month = jan,
        pages = {2937},
       adsurl = {https://ui.adsabs.harvard.edu/abs/2013ICRC...33.2937Z}
}

@ARTICLE{Fomin1994,
       author = {{Fomin}, V.~P. and others},
        title = "{New methods of atmospheric Cherenkov imaging for gamma-ray astronomy. I. The false source method}",
      journal = {Astroparticle Physics},
         year = 1994,
        month = may,
       volume = {2},
       number = {2},
        pages = {137-150},
          doi = {10.1016/0927-6505(94)90036-1},
       adsurl = {https://ui.adsabs.harvard.edu/abs/1994APh.....2..137F}
}

@ARTICLE{LiMa1983,
       author = {{Li}, T. -P. and {Ma}, Y. -Q.},
        title = "{Analysis methods for results in gamma-ray astronomy.}",
      journal = {\apj},
         year = 1983,
        month = sep,
       volume = {272},
        pages = {317-324},
          doi = {10.1086/161295},
       adsurl = {https://ui.adsabs.harvard.edu/abs/1983ApJ...272..317L}
}

@ARTICLE{Aleksic2016b,
       author = {{Aleksi{\'c}}, J. and others},
        title = "{The major upgrade of the MAGIC telescopes, Part II: A performance study using observations of the Crab Nebula}",
      journal = {Astroparticle Physics},
         year = 2016,
        month = jan,
       volume = {72},
        pages = {76-94},
          doi = {10.1016/j.astropartphys.2015.02.005},
archivePrefix = {arXiv},
       eprint = {1409.5594},
 primaryClass = {astro-ph.IM},
       adsurl = {https://ui.adsabs.harvard.edu/abs/2016APh....72...76A}
}

@ARTICLE{Mirabel1994,
       author = {{Mirabel}, I.~F. and {Rodr{\'\i}guez}, L.~F.},
        title = "{A superluminal source in the Galaxy}",
      journal = {\nat},
         year = 1994,
        month = sep,
       volume = {371},
       number = {6492},
        pages = {46-48},
          doi = {10.1038/371046a0},
       adsurl = {https://ui.adsabs.harvard.edu/abs/1994Natur.371...46M}
}

@ARTICLE{Reid2023,
       author = {{Reid}, M.~J. and {Miller-Jones}, J.~C.~A.},
        title = "{On the Distances to the X-Ray Binaries Cygnus X-3 and GRS 1915+105}",
      journal = {\apj},
         year = 2023,
        month = dec,
       volume = {959},
       number = {2},
          eid = {85},
        pages = {85},
          doi = {10.3847/1538-4357/acfe0c},
archivePrefix = {arXiv},
       eprint = {2309.15027},
 primaryClass = {astro-ph.HE},
       adsurl = {https://ui.adsabs.harvard.edu/abs/2023ApJ...959...85R}
}

@ARTICLE{Zdziarski2013,
       author = {{Zdziarski}, A.~A. and {Mikolajewska}, J. and {Belczynski}, K.},
        title = "{Cyg X-3: a low-mass black hole or a neutron star.}",
      journal = {\mnras},
         year = 2013,
        month = feb,
       volume = {429},
        pages = {L104-L108},
          doi = {10.1093/mnrasl/sls035},
archivePrefix = {arXiv},
       eprint = {1208.5455},
 primaryClass = {astro-ph.HE},
       adsurl = {https://ui.adsabs.harvard.edu/abs/2013MNRAS.429L.104Z}
}

@ARTICLE{Koljonen2017,
       author = {{Koljonen}, K.~I.~I. and {Maccarone}, T.~J.},
        title = "{Gemini/GNIRS infrared spectroscopy of the Wolf-Rayet stellar wind in Cygnus X-3}",
      journal = {\mnras},
         year = 2017,
        month = dec,
       volume = {472},
       number = {2},
        pages = {2181-2195},
          doi = {10.1093/mnras/stx2106},
archivePrefix = {arXiv},
       eprint = {1708.04050},
 primaryClass = {astro-ph.HE},
       adsurl = {https://ui.adsabs.harvard.edu/abs/2017MNRAS.472.2181K}
}

@ARTICLE{vanKerkwijk1992,
       author = {{van Kerkwijk}, M.~H. and others},
        title = "{Infrared helium emission lines from Cygnus X-3 suggesting a Wolf-Rayet star companion}",
      journal = {\nat},
         year = 1992,
        month = feb,
       volume = {355},
       number = {6362},
        pages = {703-705},
          doi = {10.1038/355703a0},
       adsurl = {https://ui.adsabs.harvard.edu/abs/1992Natur.355..703V}
}

@ARTICLE{Corbel2012,
       author = {{Corbel}, S. and others},
        title = "{A giant radio flare from Cygnus X-3 with associated {\ensuremath{\gamma}}-ray emission}",
      journal = {\mnras},
         year = 2012,
        month = apr,
       volume = {421},
       number = {4},
        pages = {2947-2955},
          doi = {10.1111/j.1365-2966.2012.20517.x},
archivePrefix = {arXiv},
       eprint = {1201.3356},
 primaryClass = {astro-ph.HE},
       adsurl = {https://ui.adsabs.harvard.edu/abs/2012MNRAS.421.2947C}
}

@ARTICLE{Fermi2009,
       author = {{Fermi LAT Collaboration} and others},
        title = "{Modulated High-Energy Gamma-Ray Emission from the Microquasar Cygnus X-3}",
      journal = {Science},
         year = 2009,
        month = dec,
       volume = {326},
       number = {5959},
        pages = {1512},
          doi = {10.1126/science.1182174},
       adsurl = {https://ui.adsabs.harvard.edu/abs/2009Sci...326.1512F}
}

@ARTICLE{Aleksic2010,
       author = {{Aleksi{\'c}}, J. and others},
        title = "{Magic Constraints on {\ensuremath{\gamma}}-ray Emission from Cygnus X-3}",
      journal = {\apj},
         year = 2010,
        month = sep,
       volume = {721},
       number = {1},
        pages = {843-855},
          doi = {10.1088/0004-637X/721/1/843},
archivePrefix = {arXiv},
       eprint = {1005.0740},
 primaryClass = {astro-ph.HE},
       adsurl = {https://ui.adsabs.harvard.edu/abs/2010ApJ...721..843A}
}

@ARTICLE{Archambault2013,
       author = {{Archambault}, S. and others},
        title = "{VERITAS Observations of the Microquasar Cygnus X-3}",
      journal = {\apj},
         year = 2013,
        month = dec,
       volume = {779},
       number = {2},
          eid = {150},
        pages = {150},
          doi = {10.1088/0004-637X/779/2/150},
archivePrefix = {arXiv},
       eprint = {1311.0919},
 primaryClass = {astro-ph.HE},
       adsurl = {https://ui.adsabs.harvard.edu/abs/2013ApJ...779..150A}
}

@ARTICLE{LHAASO2025,
       author = {{LHAASO Collaboration}},
        title = "{Ultrahigh-Energy Gamma-ray Emission Associated with Black Hole-Jet Systems}",
      journal = {National Science Review},
         year = 2025,
        month = dec,
       volume = {12},
       number = {12},
          eid = {nwaf496},
        pages = {nwaf496},
          doi = {10.1093/nsr/nwaf496},
archivePrefix = {arXiv},
       eprint = {2410.08988},
 primaryClass = {astro-ph.HE},
       adsurl = {https://ui.adsabs.harvard.edu/abs/2025NSRev..12af496L}
}

@ARTICLE{Dubus2010,
       author = {{Dubus}, G. and {Cerutti}, B. and {Henri}, G.},
        title = "{The relativistic jet of Cygnus X-3 in gamma-rays}",
      journal = {\mnras},
         year = 2010,
        month = may,
       volume = {404},
       number = {1},
        pages = {L55-L59},
          doi = {10.1111/j.1745-3933.2010.00834.x},
archivePrefix = {arXiv},
       eprint = {1002.3888},
 primaryClass = {astro-ph.HE},
       adsurl = {https://ui.adsabs.harvard.edu/abs/2010MNRAS.404L..55D}
}

@ARTICLE{Romero2017,
       author = {{Romero}, Gustavo E. and others},
        title = "{Relativistic Jets in Active Galactic Nuclei and Microquasars}",
      journal = {\ssr},
         year = 2017,
        month = jul,
       volume = {207},
       number = {1-4},
        pages = {5-61},
          doi = {10.1007/s11214-016-0328-2},
archivePrefix = {arXiv},
       eprint = {1611.09507},
 primaryClass = {astro-ph.HE},
       adsurl = {https://ui.adsabs.harvard.edu/abs/2017SSRv..207....5R}
}

@ARTICLE{Romero2003,
       author = {{Romero}, G.~E. and others},
        title = "{Hadronic gamma-ray emission from windy microquasars}",
      journal = {\aap},
         year = 2003,
        month = oct,
       volume = {410},
        pages = {L1-L4},
          doi = {10.1051/0004-6361:20031314-1},
archivePrefix = {arXiv},
       eprint = {astro-ph/0309123},
 primaryClass = {astro-ph},
       adsurl = {https://ui.adsabs.harvard.edu/abs/2003A&A...410L...1R}
}

@INPROCEEDINGS{vanHeuvel2019,
       author = {{van den Heuvel}, Edward P.~J.},
        title = "{High-Mass X-ray Binaries: progenitors of double compact objects}",
    booktitle = {High-mass X-ray Binaries: Illuminating the Passage from Massive Binaries to Merging Compact Objects},
         year = 2019,
       editor = {{Oskinova}, Lidia M. and {Bozzo}, Enrico and {Bulik}, Tomasz and {Gies}, Douglas R.},
       series = {IAU Symposium},
       volume = {346},
        month = dec,
        pages = {1-13},
          doi = {10.1017/S1743921319001315},
archivePrefix = {arXiv},
       eprint = {1901.06939},
 primaryClass = {astro-ph.HE},
       adsurl = {https://ui.adsabs.harvard.edu/abs/2019IAUS..346....1V}
}

@ARTICLE{Fender2009,
       author = {{Fender}, R.~P. and {Homan}, J. and {Belloni}, T.~M.},
        title = "{Jets from black hole X-ray binaries: testing, refining and extending empirical models for the coupling to X-rays}",
      journal = {\mnras},
         year = 2009,
        month = jul,
       volume = {396},
       number = {3},
        pages = {1370-1382},
          doi = {10.1111/j.1365-2966.2009.14841.x},
archivePrefix = {arXiv},
       eprint = {0903.5166},
 primaryClass = {astro-ph.HE},
       adsurl = {https://ui.adsabs.harvard.edu/abs/2009MNRAS.396.1370F}
}

@ARTICLE{Homan2005,
       author = {{Homan}, Jeroen and {Belloni}, Tomaso},
        title = "{The Evolution of Black Hole States}",
      journal = {\apss},
         year = 2005,
        month = nov,
       volume = {300},
       number = {1-3},
        pages = {107-117},
          doi = {10.1007/s10509-005-1197-4},
archivePrefix = {arXiv},
       eprint = {astro-ph/0412597},
 primaryClass = {astro-ph},
       adsurl = {https://ui.adsabs.harvard.edu/abs/2005Ap&SS.300..107H}
}

@ARTICLE{Remillard2006,
       author = {{Remillard}, Ronald A. and {McClintock}, Jeffrey E.},
        title = "{X-Ray Properties of Black-Hole Binaries}",
      journal = {\araa},
         year = 2006,
        month = sep,
       volume = {44},
       number = {1},
        pages = {49-92},
          doi = {10.1146/annurev.astro.44.051905.092532},
archivePrefix = {arXiv},
       eprint = {astro-ph/0606352},
 primaryClass = {astro-ph},
       adsurl = {https://ui.adsabs.harvard.edu/abs/2006ARA&A..44...49R}
}

@INCOLLECTION{McClintock2006,
       author = {{McClintock}, Jeffrey E. and {Remillard}, Ronald A.},
        title = "{Black hole binaries}",
    booktitle = {Compact stellar X-ray sources},
         year = 2006,
    publisher = {Cambridge University Press},
       editor = {{Lewin}, Walter H.~G. and {van der Klis}, Michiel},
       volume = {39},
        pages = {157-213},
          doi = {10.48550/arXiv.astro-ph/0306213},
       adsurl = {https://ui.adsabs.harvard.edu/abs/2006csxs.book..157M}
}

@ARTICLE{Fender2004,
       author = {{Fender}, R.~P. and {Belloni}, T.~M. and {Gallo}, E.},
        title = "{Towards a unified model for black hole X-ray binary jets}",
      journal = {\mnras},
         year = 2004,
        month = dec,
       volume = {355},
       number = {4},
        pages = {1105-1118},
          doi = {10.1111/j.1365-2966.2004.08384.x},
archivePrefix = {arXiv},
       eprint = {astro-ph/0409360},
 primaryClass = {astro-ph},
       adsurl = {https://ui.adsabs.harvard.edu/abs/2004MNRAS.355.1105F}
}

@ARTICLE{Koljonen2010,
       author = {{Koljonen}, K.~I.~I. and others},
        title = "{The hardness-intensity diagram of Cygnus X-3: revisiting the radio/X-ray states}",
      journal = {\mnras},
         year = 2010,
        month = jul,
       volume = {406},
       number = {1},
        pages = {307-319},
          doi = {10.1111/j.1365-2966.2010.16722.x},
archivePrefix = {arXiv},
       eprint = {1003.4351},
 primaryClass = {astro-ph.HE},
       adsurl = {https://ui.adsabs.harvard.edu/abs/2010MNRAS.406..307K}
}

@ARTICLE{Zdziarski2018,
       author = {{Zdziarski}, Andrzej A. and others},
        title = "{A comprehensive study of high-energy gamma-ray and radio emission from Cyg X-3}",
      journal = {\mnras},
         year = 2018,
        month = oct,
       volume = {479},
       number = {4},
        pages = {4399-4415},
          doi = {10.1093/mnras/sty1618},
archivePrefix = {arXiv},
       eprint = {1804.07460},
 primaryClass = {astro-ph.HE},
       adsurl = {https://ui.adsabs.harvard.edu/abs/2018MNRAS.479.4399Z}
}

@ARTICLE{atwood2009,
       author = {{Atwood}, W.~B. and others},
        title = "{The Large Area Telescope on the Fermi Gamma-Ray Space Telescope Mission}",
      journal = {\apj},
         year = 2009,
        month = jun,
       volume = {697},
       number = {2},
        pages = {1071-1102},
          doi = {10.1088/0004-637X/697/2/1071},
archivePrefix = {arXiv},
       eprint = {0902.1089},
 primaryClass = {astro-ph.IM},
       adsurl = {https://ui.adsabs.harvard.edu/abs/2009ApJ...697.1071A}
}

@PHDTHESIS{Zabalza2011,
       author = {{Zabalza}, V.},
        title = "{The keV-TeV connection in gamma-ray binaries}",
       school = {University of Barcelona, Department of Astronomy and Meteorology},
         year = 2011,
        month = may,
       adsurl = {https://ui.adsabs.harvard.edu/abs/2011PhDT.........1Z}
}

@ARTICLE{Abdo2009,
       author = {{Abdo}, A.~A. and others},
        title = "{Fermi/Large Area Telescope Bright Gamma-Ray Source List}",
      journal = {\apjs},
         year = 2009,
        month = jul,
       volume = {183},
       number = {1},
        pages = {46-66},
          doi = {10.1088/0067-0049/183/1/46},
archivePrefix = {arXiv},
       eprint = {0902.1340},
 primaryClass = {astro-ph.HE},
       adsurl = {https://ui.adsabs.harvard.edu/abs/2009ApJS..183...46A}
}

@ARTICLE{Ballet2023,
       author = {{Ballet}, J. and others},
        title = "{Fermi Large Area Telescope Fourth Source Catalog Data Release 4 (4FGL-DR4)}",
      journal = {arXiv e-prints},
         year = 2023,
        month = jul,
          eid = {arXiv:2307.12546},
        pages = {arXiv:2307.12546},
          doi = {10.48550/arXiv.2307.12546},
archivePrefix = {arXiv},
       eprint = {2307.12546},
 primaryClass = {astro-ph.HE},
       adsurl = {https://ui.adsabs.harvard.edu/abs/2023arXiv230712546B}
}

@ARTICLE{Li2018,
       author = {{Li}, K.~L. and others},
        title = "{The X-Ray Modulation of PSR J2032+4127/MT91 213 during the Periastron Passage in 2017}",
      journal = {\apj},
         year = 2018,
        month = apr,
       volume = {857},
       number = {2},
          eid = {123},
        pages = {123},
          doi = {10.3847/1538-4357/aab848},
archivePrefix = {arXiv},
       eprint = {1803.06703},
 primaryClass = {astro-ph.HE},
       adsurl = {https://ui.adsabs.harvard.edu/abs/2018ApJ...857..123L}
}

@ARTICLE{Abdo2009b,
       author = {{Abdo}, A.~A. and others},
        title = "{Detection of 16 Gamma-Ray Pulsars Through Blind Frequency Searches Using the Fermi LAT}",
      journal = {Science},
         year = 2009,
        month = aug,
       volume = {325},
       number = {5942},
        pages = {840},
          doi = {10.1126/science.1175558},
archivePrefix = {arXiv},
       eprint = {1009.0748},
 primaryClass = {astro-ph.GA},
       adsurl = {https://ui.adsabs.harvard.edu/abs/2009Sci...325..840A}
}

@ARTICLE{Ackermann2011,
       author = {{Ackermann}, M. and others},
        title = "{A Cocoon of Freshly Accelerated Cosmic Rays Detected by Fermi in the Cygnus Superbubble}",
      journal = {Science},
         year = 2011,
        month = nov,
       volume = {334},
       number = {6059},
        pages = {1103},
          doi = {10.1126/science.1210311},
       adsurl = {https://ui.adsabs.harvard.edu/abs/2011Sci...334.1103A}
}

@ARTICLE{Antokhin2019,
       author = {{Antokhin}, Igor I. and {Cherepashchuk}, Anatol M.},
        title = "{The Period Change of Cyg X-3}",
      journal = {\apj},
         year = 2019,
        month = feb,
       volume = {871},
       number = {2},
          eid = {244},
        pages = {244},
          doi = {10.3847/1538-4357/aafb38},
archivePrefix = {arXiv},
       eprint = {1807.00817},
 primaryClass = {astro-ph.HE},
       adsurl = {https://ui.adsabs.harvard.edu/abs/2019ApJ...871..244A}
}

@ARTICLE{Zdziarski2012,
       author = {{Zdziarski}, Andrzej A. and others},
        title = "{Energy-dependent orbital modulation of X-rays and constraints on emission of the jet in Cyg X-3}",
      journal = {\mnras},
         year = 2012,
        month = oct,
       volume = {426},
       number = {2},
        pages = {1031-1042},
          doi = {10.1111/j.1365-2966.2012.21635.x},
archivePrefix = {arXiv},
       eprint = {1205.4402},
 primaryClass = {astro-ph.HE},
       adsurl = {https://ui.adsabs.harvard.edu/abs/2012MNRAS.426.1031Z}
}

@ARTICLE{Nolan2012,
       author = {{Nolan}, P.~L. and others},
        title = "{Fermi Large Area Telescope Second Source Catalog}",
      journal = {\apjs},
         year = 2012,
        month = apr,
       volume = {199},
       number = {2},
          eid = {31},
        pages = {31},
          doi = {10.1088/0067-0049/199/2/31},
archivePrefix = {arXiv},
       eprint = {1108.1435},
 primaryClass = {astro-ph.HE},
       adsurl = {https://ui.adsabs.harvard.edu/abs/2012ApJS..199...31N}
}

@ARTICLE{LHAASOcocoon2024,
       author = {{LHAASO Collaboration}},
        title = "{An ultrahigh-energy {\ensuremath{\gamma}} -ray bubble powered by a super PeVatron}",
      journal = {Science Bulletin},
         year = 2024,
        month = feb,
       volume = {69},
       number = {4},
        pages = {449-457},
          doi = {10.1016/j.scib.2023.12.040},
archivePrefix = {arXiv},
       eprint = {2310.10100},
 primaryClass = {astro-ph.HE},
       adsurl = {https://ui.adsabs.harvard.edu/abs/2024SciBu..69..449L}
}

@ARTICLE{tevj2032,
       author = {{Aharonian}, F. and others},
        title = "{An unidentified TeV source in the vicinity of Cygnus OB2}",
      journal = {\aap},
         year = 2002,
        month = oct,
       volume = {393},
        pages = {L37-L40},
          doi = {10.1051/0004-6361:20021171},
archivePrefix = {arXiv},
       eprint = {astro-ph/0207528},
 primaryClass = {astro-ph},
       adsurl = {https://ui.adsabs.harvard.edu/abs/2002A&A...393L..37A}
}

@ARTICLE{Cao2020,
       author = {{Cao}, Xinwu and {Zdziarski}, Andrzej A.},
        title = "{Jets in the soft state in Cyg X-3 caused by advection of the donor magnetic field and unification with low-mass X-ray binaries}",
      journal = {\mnras},
         year = 2020,
        month = feb,
       volume = {492},
       number = {1},
        pages = {223-231},
          doi = {10.1093/mnras/stz3447},
archivePrefix = {arXiv},
       eprint = {1910.01377},
 primaryClass = {astro-ph.HE},
       adsurl = {https://ui.adsabs.harvard.edu/abs/2020MNRAS.492..223C}
}

@ARTICLE{Rolke2005,
       author = {{Rolke}, Wolfgang A. and {L{\'o}pez}, Angel M. and {Conrad}, Jan},
        title = "{Limits and confidence intervals in the presence of nuisance parameters}",
      journal = {Nuclear Instruments and Methods in Physics Research A},
         year = 2005,
        month = oct,
       volume = {551},
       number = {2-3},
        pages = {493-503},
          doi = {10.1016/j.nima.2005.05.068},
archivePrefix = {arXiv},
       eprint = {physics/0403059},
 primaryClass = {physics.data-an},
       adsurl = {https://ui.adsabs.harvard.edu/abs/2005NIMPA.551..493R}
}

@ARTICLE{Aleksic2012,
       author = {{Aleksi{\'c}}, J. and others},
        title = "{Performance of the MAGIC stereo system obtained with Crab Nebula data}",
      journal = {Astroparticle Physics},
         year = 2012,
        month = feb,
       volume = {35},
       number = {7},
        pages = {435-448},
          doi = {10.1016/j.astropartphys.2011.11.007},
archivePrefix = {arXiv},
       eprint = {1108.1477},
 primaryClass = {astro-ph.IM},
       adsurl = {https://ui.adsabs.harvard.edu/abs/2012APh....35..435A}
}

@ARTICLE{V4641Sgr,
       author = {{Alfaro}, R. and others},
        title = "{Ultra-high-energy gamma-ray bubble around microquasar V4641 Sgr}",
      journal = {\nat},
         year = 2024,
        month = oct,
       volume = {634},
       number = {8034},
        pages = {557-560},
          doi = {10.1038/s41586-024-07995-9},
archivePrefix = {arXiv},
       eprint = {2410.16117},
 primaryClass = {astro-ph.HE},
       adsurl = {https://ui.adsabs.harvard.edu/abs/2024Natur.634..557A}
}

@ARTICLE{Dmytriiev2024,
       author = {{Dmytriiev}, Anton and others},
        title = "{Two Models for the Orbital Modulation of Gamma Rays in Cyg X-3}",
      journal = {\apj},
         year = 2024,
        month = sep,
       volume = {972},
       number = {1},
          eid = {85},
        pages = {85},
          doi = {10.3847/1538-4357/ad6440},
archivePrefix = {arXiv},
       eprint = {2405.09154},
 primaryClass = {astro-ph.HE},
       adsurl = {https://ui.adsabs.harvard.edu/abs/2024ApJ...972...85D}
}

@ARTICLE{Gould1967,
       author = {{Gould}, Robert J. and {Schr{\'e}der}, G{\'e}rard P.},
        title = "{Pair Production in Photon-Photon Collisions}",
      journal = {Physical Review},
         year = 1967,
        month = mar,
       volume = {155},
       number = {5},
        pages = {1404-1407},
          doi = {10.1103/PhysRev.155.1404},
       adsurl = {https://ui.adsabs.harvard.edu/abs/1967PhRv..155.1404G}
}

@ARTICLE{Egron2017,
       author = {{Egron}, E. and others},
        title = "{Single-dish and VLBI observations of Cygnus X-3 during the 2016 giant flare episode}",
      journal = {\mnras},
         year = 2017,
        month = nov,
       volume = {471},
       number = {3},
        pages = {2703-2714},
          doi = {10.1093/mnras/stx1730},
archivePrefix = {arXiv},
       eprint = {1707.03761},
 primaryClass = {astro-ph.HE},
       adsurl = {https://ui.adsabs.harvard.edu/abs/2017MNRAS.471.2703E}
}

@ARTICLE{Bosch2009,
       author = {{Bosch-Ramon}, Valent{\'\i} and {Khangulyan}, Dmitry},
        title = "{Understanding the Very-High Emission from Microquasars}",
      journal = {International Journal of Modern Physics D},
         year = 2009,
        month = jan,
       volume = {18},
       number = {3},
        pages = {347-387},
          doi = {10.1142/S0218271809014601},
archivePrefix = {arXiv},
       eprint = {0805.4123},
 primaryClass = {astro-ph},
       adsurl = {https://ui.adsabs.harvard.edu/abs/2009IJMPD..18..347B}
}

@ARTICLE{agile2012,
       author = {{Bulgarelli}, A. and others},
        title = "{AGILE detection of Cygnus X-3 {\ensuremath{\gamma}}-ray active states during the period mid-2009/mid-2010}",
      journal = {\aap},
         year = 2012,
        month = feb,
       volume = {538},
          eid = {A63},
        pages = {A63},
          doi = {10.1051/0004-6361/201016129},
archivePrefix = {arXiv},
       eprint = {1111.4960},
 primaryClass = {astro-ph.HE},
       adsurl = {https://ui.adsabs.harvard.edu/abs/2012A&A...538A..63B}
}

@ARTICLE{ovro2011,
       author = {{Richards}, Joseph L. and others},
        title = "{Blazars in the Fermi Era: The OVRO 40 m Telescope Monitoring Program}",
      journal = {\apjs},
         year = 2011,
        month = jun,
       volume = {194},
       number = {2},
          eid = {29},
        pages = {29},
          doi = {10.1088/0067-0049/194/2/29},
archivePrefix = {arXiv},
       eprint = {1011.3111},
 primaryClass = {astro-ph.CO},
       adsurl = {https://ui.adsabs.harvard.edu/abs/2011ApJS..194...29R}
}

@ARTICLE{maxi2009,
       author = {{Matsuoka}, Masaru and others},
        title = "{The MAXI Mission on the ISS: Science and Instruments for Monitoring All-Sky X-Ray Images}",
      journal = {\pasj},
         year = 2009,
        month = oct,
       volume = {61},
        pages = {999},
          doi = {10.1093/pasj/61.5.999},
archivePrefix = {arXiv},
       eprint = {0906.0631},
 primaryClass = {astro-ph.IM},
       adsurl = {https://ui.adsabs.harvard.edu/abs/2009PASJ...61..999M}
}

@ARTICLE{swift2013,
       author = {{Krimm}, H.~A. and others},
        title = "{The Swift/BAT Hard X-Ray Transient Monitor}",
      journal = {\apjs},
         year = 2013,
        month = nov,
       volume = {209},
       number = {1},
          eid = {14},
        pages = {14},
          doi = {10.1088/0067-0049/209/1/14},
archivePrefix = {arXiv},
       eprint = {1309.0755},
 primaryClass = {astro-ph.HE},
       adsurl = {https://ui.adsabs.harvard.edu/abs/2013ApJS..209...14K}
}

@ARTICLE{Khangulyan2014,
       author = {{Khangulyan}, D. and {Aharonian}, F.~A. and {Kelner}, S.~R.},
        title = "{Simple Analytical Approximations for Treatment of Inverse Compton Scattering of Relativistic Electrons in the Blackbody Radiation Field}",
      journal = {\apj},
         year = 2014,
        month = mar,
       volume = {783},
       number = {2},
          eid = {100},
        pages = {100},
          doi = {10.1088/0004-637X/783/2/100},
archivePrefix = {arXiv},
       eprint = {1310.7971},
 primaryClass = {astro-ph.HE},
       adsurl = {https://ui.adsabs.harvard.edu/abs/2014ApJ...783..100K}
}

@ARTICLE{Khangulyan2018,
       author = {{Khangulyan}, Dmitry and {Bosch-Ramon}, Valent{\'\i} and {Uchiyama}, Yasunobu},
        title = "{Inverse Compton emission from relativistic jets in binary systems}",
      journal = {\mnras},
         year = 2018,
        month = dec,
       volume = {481},
       number = {2},
        pages = {1455-1468},
          doi = {10.1093/mnras/sty2356},
archivePrefix = {arXiv},
       eprint = {1808.09628},
 primaryClass = {astro-ph.HE},
       adsurl = {https://ui.adsabs.harvard.edu/abs/2018MNRAS.481.1455K}
}

@ARTICLE{Bednarek2010,
       author = {{Bednarek}, W.},
        title = "{On the possibility of sub-TeV -ray emission from CygX-3}",
      journal = {\mnras},
         year = 2010,
        month = jul,
       volume = {406},
       number = {1},
        pages = {689-700},
          doi = {10.1111/j.1365-2966.2010.16721.x},
archivePrefix = {arXiv},
       eprint = {1002.1563},
 primaryClass = {astro-ph.HE},
       adsurl = {https://ui.adsabs.harvard.edu/abs/2010MNRAS.406..689B}
}

@ARTICLE{Bednarek2005,
       author = {{Bednarek}, W.},
        title = "{GeV gamma-rays and TeV neutrinos from very massive compact binary systems: the case of WR 20a}",
      journal = {\mnras},
         year = 2005,
        month = oct,
       volume = {363},
       number = {1},
        pages = {L46-L50},
          doi = {10.1111/j.1745-3933.2005.00081.x},
archivePrefix = {arXiv},
       eprint = {astro-ph/0507565},
 primaryClass = {astro-ph},
       adsurl = {https://ui.adsabs.harvard.edu/abs/2005MNRAS.363L..46B}
}

@ARTICLE{Levinson2001,
       author = {{Levinson}, Amir and {Waxman}, Eli},
        title = "{Probing Microquasars with TeV Neutrinos}",
      journal = {\prl},
         year = 2001,
        month = oct,
       volume = {87},
       number = {17},
          eid = {171101},
        pages = {171101},
          doi = {10.1103/PhysRevLett.87.171101},
archivePrefix = {arXiv},
       eprint = {hep-ph/0106102},
 primaryClass = {hep-ph},
       adsurl = {https://ui.adsabs.harvard.edu/abs/2001PhRvL..87q1101L}
}

@ARTICLE{cta2025,
       author = {{Abe}, K. and others},
        title = "{Galactic transient sources with the Cherenkov Telescope Array Observatory}",
      journal = {\mnras},
         year = 2025,
        month = jun,
       volume = {540},
       number = {1},
        pages = {205-238},
          doi = {10.1093/mnras/staf655},
archivePrefix = {arXiv},
       eprint = {2405.04469},
 primaryClass = {astro-ph.HE},
       adsurl = {https://ui.adsabs.harvard.edu/abs/2025MNRAS.540..205A}
}

@ARTICLE{Marti1992,
       author = {{Marti}, J. and {Paredes}, J.~M. and {Estalella}, R.},
        title = "{Modelling Cygnus X-3 radio outbursts : particle injection into twin jets.}",
      journal = {\aap},
         year = 1992,
        month = may,
       volume = {258},
        pages = {309},
       adsurl = {https://ui.adsabs.harvard.edu/abs/1992A&A...258..309M}
}

@ARTICLE{Marti2001,
       author = {{Mart{\'\i}}, J. and {Paredes}, J.~M. and {Peracaula}, M.},
        title = "{Development of a two-sided relativistic jet in Cygnus X-3}",
      journal = {\aap},
         year = 2001,
        month = aug,
       volume = {375},
        pages = {476-484},
          doi = {10.1051/0004-6361:20010907},
       adsurl = {https://ui.adsabs.harvard.edu/abs/2001A&A...375..476M}
}

@ARTICLE{LHAASO2026CX3,
       author = {{LHAASO Collaboration}},
        title = "{Cygnus X-3: A variable petaelectronvolt gamma-ray source}",
      journal = {arXiv e-prints},
         year = 2025,
        month = dec,
          eid = {arXiv:2512.16638},
        pages = {arXiv:2512.16638},
          doi = {10.48550/arXiv.2512.16638},
archivePrefix = {arXiv},
       eprint = {2512.16638},
 primaryClass = {astro-ph.HE},
       adsurl = {https://ui.adsabs.harvard.edu/abs/2025arXiv251216638T}
}

@ARTICLE{AGILE2009,
       author = {{Tavani}, M. and others},
        title = "{Extreme particle acceleration in the microquasar CygnusX-3}",
      journal = {\nat},
         year = 2009,
        month = dec,
       volume = {462},
       number = {7273},
        pages = {620-623},
          doi = {10.1038/nature08578},
archivePrefix = {arXiv},
       eprint = {0910.5344},
 primaryClass = {astro-ph.HE},
       adsurl = {https://ui.adsabs.harvard.edu/abs/2009Natur.462..620T}
}

@ARTICLE{Zhang2026,
       author = {{Zhang}, Xing-Fu and others},
        title = "{Constraining the PeV gamma-ray emission zone of Cygnus X-3 with contemporaneous GeV timing and spectral observations}",
      journal = {arXiv e-prints},
         year = 2026,
        month = apr,
          eid = {arXiv:2604.04768},
        pages = {arXiv:2604.04768},
          doi = {10.48550/arXiv.2604.04768},
archivePrefix = {arXiv},
       eprint = {2604.04768},
 primaryClass = {astro-ph.HE},
       adsurl = {https://ui.adsabs.harvard.edu/abs/2026arXiv260404768Z}
}

@ARTICLE{Koljonen2018,
       author = {{Koljonen}, K.~I.~I. and others},
        title = "{The hypersoft state of Cygnus X-3. A key to jet quenching in X-ray binaries?}",
      journal = {\aap},
         year = 2018,
        month = apr,
       volume = {612},
          eid = {A27},
        pages = {A27},
          doi = {10.1051/0004-6361/201732284},
archivePrefix = {arXiv},
       eprint = {1712.07933},
 primaryClass = {astro-ph.HE},
       adsurl = {https://ui.adsabs.harvard.edu/abs/2018A&A...612A..27K}
}
